# Terrestrial modification of the Ivuna meteorite and a reassessment of the chemical composition of the CI type specimen

A. J. King[1†], K. J. H. Phillips[1*], S. Strekopytov[2,‡], C. Vita-Finzi[1*] and S. S. Russell[1]

[1]*Planetary Materials Group, Department of Earth Sciences, Natural History Museum, Cromwell Road, London SW7 5BD, U.K.*

[2]*Imaging and Analysis Centre, Natural History Museum, Cromwell Road, London SW7 5BD, U.K.*

†Corresponding Author: A. J. King (a.king@nhm.ac.uk), Phone: +44 (0)20 7942 6979

‡Current Address: *Inorganic Analysis, LGC, Queens Road, Teddington, Middlesex, TW11 0LY, U. K.*

*Scientific Associate

**The rare CI carbonaceous chondrites are the most aqueously altered and chemically primitive meteorites but due to their porous nature and high abundance of volatile elements are susceptible to terrestrial weathering. The Ivuna meteorite, type specimen for the CI chondrites, is the largest twentieth-century CI fall and probably the CI chondrite least affected by terrestrial alteration that is available for study. The main mass of Ivuna (BM2008 M1) has been stored in a nitrogen atmosphere at least since its arrival at the Natural History Museum (NHM), London, in 2008 (70 years after its fall) and could be considered the most pristine CI chondrite stone. We report the mineralogy, petrography and bulk elemental composition of BM2008 M1 and a second Ivuna stone (BM1996 M4) stored in air within wooden cabinets. We find that both Ivuna stones are breccias consisting of multiple rounded, phyllosilicate-rich clasts that formed through aqueous alteration followed by impact processing. A polished thin section of BM2008 M1 analysed immediately after preparation was found to contain sulphate-bearing veins that**

**formed when primary sulphides reacted with oxygen and atmospheric water. A section of BM1996 M4 lacked veins but had sulphate grains on the surface that formed in ≤6 years, ~3 times faster than previous reports for CI chondrite sections. Differences in the extent of terrestrial alteration recorded by BM2008 M1 and BM1996 M4 probably reflect variations in the post-recovery curation history of the stones prior to entering the NHM collection, and indicate that where possible pristine samples of hydrated carbonaceous should be kept out of the terrestrial environment in a stable environment to avoid modification. The bulk elemental composition of the two Ivuna stones show some variability due to their heterogeneous nature but in general are similar to previous analyses of CI chondrites. We combine our elemental abundances with literature values to calculate a new average composition for the Ivuna meteorite, which we find is in good agreement with existing compilations of element compositions in the CI chondrites and the most recent solar photospheric abundances.**

## 1. Introduction

Primitive chondrite meteorites provide a direct opportunity to study the materials, processes and events that formed our solar system ~4.6-Gyr ago. However, a major challenge in using meteorites to understand the evolution of the solar system is discriminating between the features created on asteroid parent bodies and those induced by terrestrial alteration. Upon entering the Earth's atmosphere, meteorites are immediately susceptible to modification by oxidation, hydration, and carbonation, or the addition of terrestrial contaminants. Most meteorites are finds recovered from hot and cold deserts and exposed to terrestrial weathering for several thousands to millions of years (e.g. Bland et al. 2006). A much smaller number of meteorites are witnessed falls that were recovered soon (hours to years) after landing on Earth but in some cases still show signs of terrestrial weathering (e.g. Bland et al. 1998a; Lee and Bland 2004; Walker et al. 2018).

Typical mineralogical changes associated with terrestrial weathering of meteorites include oxidation of metal, sulphides, and silicates into "rusts" (Buddhue 1957; Gibson and Bogard 1978; Velbel and Gooding 1988; Ikeda and Kojima 1991; Wlotzka 1993; Bland et al. 1998a; Lee and Bland 2004), the appearance of white carbonate and sulphate evaporites on surfaces and filling interior pore spaces and veins (Marvin 1980; Jull et al. 1988; Velbel 1988; Velbel et al. 1991; Barrat et al. 1998; Gounelle and Zolensky 2001; Lee and Bland 2004), and the transformation of anhydrous primary phases into hydrous clay minerals (Gooding 1986; Zolensky and Gooding 1986; Lee and Bland 2004). Terrestrial weathering can also lead to significant mobilization, redistribution, loss and addition of major, minor, and trace elements (Gibson and Bogard 1978; Velbel et al. 1991; Bland et al. 1998b; Stelzner et al. 1999; Barrat et al. 2001, 2003; Crozaz and Wadhwa 2001; Crozaz et al. 2003; Hezel et al. 2011; Walker et al. 2018). The effects of physical and chemical weathering have been observed to occur in as little as a few hours even for those meteorites collected rapidly after the fall and stored under laboratory conditions (Gibson and Bogard 1978; Jull et al. 1988; Bland et al. 1998a; Barrat et al. 1999; Lee and Bland 2004; Velbel 2014; Walker et al. 2018).

To date, many studies have quantified the effects of terrestrial weathering on the widely available ordinary chondrites (e.g. Gibson and Bogard 1978; Jull et al. 1988; Ikeda and Kojima 1991; Velbel et al. 1991; Wlotzka 1993; Bland et al. 1998a, 1998b; Stelzner et al. 2011; Velbel 2014) but few have investigated the carbonaceous chondrite meteorites (e.g. Zolensky and Gooding 1986; Gounelle and Zolensky 2001; Velbel and Palmer 2011; Walker et al. 2018; King et al. 2019a). Such studies are crucial, however, as carbonaceous chondrites are scientifically valuable – as pristine samples of the early solar system and potential sources of water and organics to the terrestrial planets – but relatively scarce meteorites that due to their fragile, porous and chemically volatile nature are highly vulnerable to terrestrial alteration. Furthermore, the scientific goals of sample-return missions such as JAXA's Hayabusa-2 and NASA's OSIRIS-REx, both of which aim to collect samples from hydrated carbonaceous asteroids (Kitazato et al. 2019; Hamilton et al. 2019) rely upon developing suitable curation and analytical protocols that facilitate thorough investigation of the returned materials.

The small family of CI carbonaceous chondrites consists of only five meteorite falls: Alais (fell 1806; total mass ~6 kg), Orgueil (1864; ~14 kg), Tonk (1911; 7.7 g), Ivuna (1938; ~0.7 kg) and Revelstoke (1965; 1 g). We note that in the *Meteoritical Bulletin* a further four Antarctic finds are classified as CIs but may actually be members of a distinct carbonaceous chondrite group (King et al. 2019b). Of the falls only Alais, Orgueil and Ivuna are large enough for detailed analysis. The CI chondrites contain abundant phyllosilicates (>80 vol%), lack obvious chondrules and calcium-aluminium-rich inclusions (CAIs), and are some of the most hydrated extra-terrestrial materials available for study (Tomeoka and Buseck 1988; Tomeoka 1990; Zolensky et al. 1993; Endress and Bischoff 1996; Bullock et al. 2005; Morlok et al. 2006; King et al. 2015a, 2015b). Having formed through aqueous alteration on water-rich ancient asteroids, or possibly comets (Gounelle et al. 2006; Berger et al. 2011), they also have bulk chemical compositions that are believed to be like those in the infant solar system (Anders and Grevesse 1989; Lodders et al. 2009). The composition of the early solar system is important in that differences from measured present photospheric composition provide information about the processing of elements in the Sun if these are convected down to layers where they are altered by nuclear reactions. The elemental abundances of CI chondrites are similar to those of the solar photosphere with a few notable exceptions. These include noble gases and volatile elements such as H, C, N, and O, which are depleted in meteorites (Anders and Grevesse 1989; Lodders 2003), while the photospheric abundance of Li is smaller by a factor of ~150 compared with its meteoritic abundance probably because solar Li has undergone nuclear processes that have partly destroyed it over the ~4.6-Gyr lifetime of the Sun (Lodders 2003; Asplund et al. 2009).

Previous studies have shown that the fine-grained CI chondrites react easily with the terrestrial atmosphere and may experience mineralogical and chemical changes even in samples carefully curated in museum collections since they fell (Berzelius 1834; Zolensky 1999; Gounelle and Zolensky 2001). The Ivuna meteorite, the type specimen of the CI chondrites, fell near the western shore of Lake Rukwa in Tanzania on 16 December 1938 at 17.30 local time. Three stones were seen to fall, only one of which was recovered the following day, supposedly from the branches of a tree (Oates 1941). The recovered stone was delivered to the Geological Survey of Tanzania office in Dodoma where it was described as being "smooth and rounded save for one corner which appeared to have been recently broken due to careless handling" and with a "bituminous or coaly appearance" (Oates 1941). With a total known weight of ~0.7 kg, Ivuna is smaller than the Alais or Orgueil meteorites, but as the most recent fall of appreciable size it is probably the least terrestrially modified example of the CI group available for rigorous study.

The Natural History Museum (NHM), London, has three pieces of Ivuna: BM1991 M5 (~0.6 g), BM1996 M4 (~2.2 g), and the main mass BM2008 M1 (~200 g). The BM2008 M1 sample, by far the largest known fragment of the meteorite, was purchased from a private collector in 2008. Its exact history cannot be verified but, other than for a brief period during sampling, it has been stored in an inert (nitrogen) atmosphere within a sealed glass case at least since its arrival at the NHM, and is believed to have been kept under similar conditions for a significant period of its residence on Earth. To the best of our knowledge no other stones of Ivuna are curated in such a way, so it might be presumed that BM2008 M1 is in as near a pristine state as possible for any CI chondrite sample.

The NHM Ivuna stone BM2008 M1 offers a unique opportunity to evaluate the influence of terrestrial weathering and laboratory curation on hydrated carbonaceous chondrites, which is an important step towards interpreting their geological history and developing appropriate facilities for future meteorite falls and returned extra-terrestrial samples. Here we report detailed mineralogical and petrographic observations of BM2008 M1 and compare them to a second NHM Ivuna stone, BM1996 M4, which has been stored either in air or wrapped in fluoropolymer film within wooden cabinets. Our aim was to characterise weathering products in Ivuna and examine whether the nitrogen storage of BM2008 M1 made it less susceptible to, or suppressed the effects of, terrestrial alteration. In addition, despite being the type specimen for the CI chondrites Ivuna has far fewer chemical data in the literature associated with it than Alais or Orgueil due to its lower available mass. We therefore also present bulk major, minor and trace element abundances for both BM2008 M1 and BM1996 M4, and combine our new data with published values to determine a new average bulk composition for the Ivuna meteorite.

## 2. Experimental

### 2.1 *Samples*

As the main mass (~200 g) of Ivuna, the BM2008 M1 sample curated at the NHM merits a brief description. It is a rounded, friable, porous, and partially fusion crusted stone, ~5 cm in maximum dimension (Fig. 1). The fusion crust is very dark brown to black, and contains sparsely distributed vesicles visible on the surface. White efflorescence and mm-sized bright flecks can be seen close to the fusion crust but otherwise the interior is very fine-grained, jet black and featureless.

Polished thin sections of the Ivuna stones BM2008 M1 (P16384) and BM1996 M4 (P12628) were prepared in the same way using ethylene glycol (ethane-1,2-diol) rather than water to minimise potential alteration during the section production. The fragile samples were impregnated with epoxy resin to stabilise them during preparation of the thin sections. The BM2008 M1 section was characterised using a scanning electron microscope (SEM) (see Section 2.2) immediately after preparation to avoid degradation of the sample, whereas the BM1996 M4 section was studied six years after being made.

In addition to the thin sections, we also selected chips (~40 mg) of BM2008 M1 and BM1996 M4 that were free of fusion crust and to the naked eye appeared not to contain white efflorescence or other obvious signs of terrestrial weathering. The bulk chemical composition of each chip was determined using inductively coupled plasma optical emission spectroscopy (ICP-OES) and inductively coupled plasma mass spectrometry (ICP-MS) (see Section 2.3).

### 2.2 *Electron Microscopy*

Thin sections were carbon-coated and initially analysed using a Zeiss EVO 15LS analytical SEM with an Oxford Instruments X-Max80 energy-dispersive X-ray silicon drift detector (EDS) at the NHM. The EDS system was calibrated using elemental cobalt standard and a Kakanui augite mineral standard, and secondary electron (SE) and backscattered electron (BSE) images and X-ray element maps were obtained at ~1 µm spatial resolution using an acceleration voltage of 20 kV and a beam current of 3 nA. Further SE/BSE images and X-ray element maps of the thin sections were later acquired using an FEI Quanta 650 FEG SEM equipped with a high sensitivity Bruker Flat Quad 5060F EDS using acceleration voltages of 6 kV and 9 kV.

### 2.3 *Chemical Analysis*

The major, minor, and trace element compositions of the Ivuna samples BM1996 M4 and BM2008 M1 were determined using ICP-OES and ICP-MS at the NHM following analytical procedures for bulk-rock chemical analysis similar to those described in detail by Calzada-Diaz et al. (2017). Each chip was separately powdered and homogenised in a clean laboratory (class 1000) using an agate mortar and pestle.

Approximately 20 mg of each powdered sample was pre-treated with 1 mL of concentrated nitric acid ($HNO_3$) and fused with 120 mg of lithium metaborate ($LiBO_2$) in a Pt/Au crucible. The resulting flux was dissolved in 0.64M $HNO_3$ before major and minor element abundances were measured in triplicate using a Thermo iCap Duo ICP-OES instrument. The instrument was calibrated using a range of certified reference materials (CRMs) prepared in the same way and the standards DTS-2B (dunite) and JB-1 (basalt) were analysed as unknown samples to provide quality control.

The remaining ~20 mg of each powdered sample was digested in hydrofluoric acid (HF) + perchloric acid ($HClO_4$) + $HNO_3$ before trace element abundances were measured in duplicate using an Agilent 7700x ICP-MS instrument. Indium ($^{115}In$) at a concentration of 1 $\mu g\ g^{-1}$ was used as an internal standard. The relative contribution from the naturally occurring indium in the meteorite samples was assumed to be <5% (Fouché and Smales 1967); this was taken into account in calculating the concentrations. Non-isobaric interferences were reduced by tuning $CeO^+/Ce^+$ and $Ba^{++}/Ba^+$ to <1%; for Eu, Gd, Tb, Hf, Ta, and W, corrections were applied for polyatomic interferences from Ba, (Ce+Pr), Nd, Dy, Ho, and Er, respectively (Strekopytov and Dubinin 1997; Ferrat et al. 2012). The accuracy of the trace element analyses was monitored using the CRMs BCR-2 (basalt), JLs-1 (limestone), and SY-2 (syenite).

## 3. Results

### 3.1 *Petrography*

Figure 2 shows a BSE image of a thin section prepared from the Ivuna stone BM1996 M4. The section is a breccia containing agglomerated clasts with different levels of brightness that relate to differences in the overall Fe content (i.e. Fe-rich areas appear brighter in BSE images). The clasts are typically ~1 – 3 mm in diameter, have rounded shapes, and can also be easily distinguished by variations in the abundance of potassium (Fig. 3). The clasts are composed of a fine-grained, fibrous, Fe-rich phyllosilicate mass (Fig. 4) in which grains of carbonates, sulphides and magnetite are embedded. The grain size (<1 µm) of the phyllosilicates is smaller than the spatial resolution of our analyses so quantitative phase identification is not possible, but it is probably a mixture of serpentine and saponite, as has previously been reported in CI chondrites (Tomeoka and Buseck 1988; Tomeoka 1990; Brearley 1992; Zolensky et al. 1993; King et al. 2015b). Carbonates are present as irregularly shaped grains up to 100 µm across and are located within the matrix and not along cracks. The sulphides are laths, hexagons, or irregular-shaped grains up to 50 µm across and are typically pyrrhotite. We did not notice any grains of pentlandite and cubanite, which have previously been found in CI chondrites (Bullock et al. 2005; Berger et al. 2011), but we did not specifically look for them. Magnetite is typically present as 1 – 10 µm rounded grains that are either embedded in the matrix as isolated features or occur in clusters. It is apparent from the BSE images and X-ray element maps that the abundance of carbonates, sulphides and magnetite differs between the clasts.

Sulphate grains are clearly visible on the Ivuna BM1996 M4 thin section. They have delicate structures and grow out of the polished surface, on both phyllosilicates and sulphides, suggesting that they formed *in-situ* in the years following the original sample preparation (Fig. 5). The thin section also contains numerous cracks and fractures (Fig. 2) that are often enriched in carbon (Fig. 6). This carbon is probably terrestrial contamination from the epoxy resin used to stabilise the sample during preparation of the thin section, although there are also pockets of carbonaceous-rich material in the matrix which may be indigenous to the meteorite. The cracks and fractures do not show significant enrichments in any other elements.

Figure 7 shows that the Ivuna stone BM2008 M1 is a breccia consisting of agglomerated mm-sized rounded clasts of varying Fe content. The dominant mineralogy of the clasts is a fine-grained Fe-rich mixture of serpentine and saponite. Grains (1 to over 100 micrometers in size) of carbonates, sulphides and magnetite are embedded within the phyllosilicate groundmass and their abundances vary between the clasts (Fig. 8). Cracks and fractures are abundant in the thin section and typically enriched in Na and S (Fig. 9). Thus, the only obvious difference between the studied thin sections of two Ivuna stones (BM1996 M4 and BM2008 M1) is that the section of BM1996 M4 does not show an enrichment of Na and S along cracks and fractures.

### 3.2 *Chemical Analysis*

The bulk major, minor, and trace element compositions of the NHM Ivuna stones BM1996 M4 and BM2008 M1 are given in Table 1. Also given are the elemental abundances reported by Barrat et al. (2012), which is the most complete dataset for a single Ivuna sample available in the literature to date.

For the major element abundances, there is good agreement between BM1996 M4 and BM2008 M1 except for Na, which is 1.4× more abundant in BM2008 M1 (Fig. 10a). The REEs are all more abundant in BM1996 M4 by a factor of 1.1 – 1.6, with the difference being largest for the heavy REEs (Fig. 10b). For the trace elements, notable differences between BM1996 M4 and BM2008 M1 include K, the abundance of which is 1.6× higher in BM2008 M1, and Y, Zr, Hf, Ta, Tl, Th and U, which are all a factor of 1.2 – 3.5 more abundant in BM1996 M4 (Fig. 10c).

Comparison to the Ivuna sample measured by Barrat et al. (2012) shows that both BM1996 M4 and BM2008 M1 contain more Al (~1.2×) and less P (~1.3×), while BM2008 M1 contains a factor of 1.3 more Na (Fig. 10a). There is reasonable agreement between the REE abundances of BM2008 M1 and the Barrat et al. (2012) sample, but Ho – Lu are a factor of 1.2 – 1.6 higher in BM1996 M4 relative to the Barrat et al. (2012) values (Fig. 10b). The abundance of K in the Barrat et al. (2012) sample falls between the NHM Ivuna stones (1.3× higher than BM1996 M4 and 1.2× lower than BM2008 M1). Other trace elements are generally in good agreement although Zr (4×) and Ta (2.6×) are more abundant in BM1996 M4 than the Barrat et al. (2012) sample, while the abundance of Y (1.4×) is lower in BM2008 M1. The abundances of Ba, Hf, W, Pb, Th, U are all higher by a factor of 1.2 – 3.9× in BM1996 M4 and BM2008 M1 relative to the study of Barrat et al. (2012) (Fig. 10c).

### 3.3 *Average Chemical Composition of Ivuna*

We have used previously published analytical data (see references in Table 1 footnote) and our own measurements to calculate an average composition for Ivuna, the type specimen for the CI carbonaceous chondrite meteorites. In order to calculate the average we performed Grubbs' test (Grubbs 1969) to detect any significant outliers resulting from sample or analytical biases in the various studies. Grubbs' test strictly works for a hypothesis that only one outlier may be present, but it is reasonably accurate for a very small number ($n$) of replicate analyses ($n \leq 9$), as is the case here.

#### 3.3.1 *Major and Minor Elements*

In addition to Barrat et al. (2012), the contents of the major and minor oxides – elements present in Ivuna with >0.08% mass concentration when re-calculated to an oxide form – have previously been reported by Wiik (1956) and Wolf and Palme (2001), the former being the first analysis of the chemical composition of Ivuna. Wolf and Palme (2001) presented the abundances of all major oxides ($Al_2O_3$, CaO, FeO, MgO, NiO, $SiO_2$), but our work is the first to report the abundances of all the major and minor oxides in Ivuna in the same study. Wiik (1956) did not report the Fe abundance, while Wolf and Palme (2001) did not report the Na abundance, and Barrat et al. (2012) did not report Si abundance. Grubbs' test showed that the abundances of Ca and P reported by Wiik (1956), and the abundance of K reported by Kaushal and Wetherill (1970) are significant outliers, so these values were excluded from calculating the average chemical composition of Ivuna. As can be seen from Table 1, the spread of Na, P and K values is large (relative standard deviation (RSD) = >10%) even after removing significant outliers. All available analytical data (*n* = 5 – 11) were used in calculating the average contents of the remaining major and minor elements.

### 3.3.2 *Rare Earth Elements*

Along with Barrat et al. (2012), REE abundances (La – Lu, note Sc and Y are listed as Trace Elements in Table 1) in Ivuna have previously been reported by Schmitt et al. (1964), Rocholl and Jochum (1993), Pourmand et al. (2012), and Braukmüller et al. (2018). The REE abundance patterns given by Schmitt et al. (1964) and Rocholl and Jochum (1993) are erratic, probably due to insufficient precision of the analytical technique used in those studies, and those La – Lu datasets were excluded in calculating the average REE abundances of Ivuna. Similarly, we find that the heavier REE (Dy – Lu) abundances in BM1996 M4 are all outliers according to Grubbs' test and these were omitted from the calculated average.

The REE abundances in BM2008 M1 were not found to be anomalous from Grubbs' test, although the Y abundance is an outlier compared with published data (Schmitt et al. 1964; Rocholl and Jochum 1993; Barrat et al. 2012; Pourmand et al. 2012; Braukmüller et al. 2018) and the BM1996 M4 analysis, so our average Y concentration for Ivuna excludes this value. No significant outliers were found for Sc and all seven available values including the two determined in this study were used to calculate the average Sc content in Ivuna.

### 3.3.3 *Trace Elements*

Barrat et al. (2012) is the most complete source of trace element data for Ivuna, although it does not include the abundances of Mo, Ag, Cd, Sn, Sb, and Tl. However, these (and other) trace elements were reported by Greenland (1967), Case et al. (1973), Krähenbühl et al. (1973), Ebihara et al. (1982), Rocholl and Jochum (1993), and Braukmüller et al. (2018). None of the trace element abundances in the literature were found to be outliers according to Grubbs' test except for the Cd and Ag measurements of Greenland (1967). In contrast, the abundances of Zr, Hf, Ta, Th, and U in the BM1996 M4 sample are significant outliers and were not used in the average Ivuna chemical composition.

## 4. Discussion

### 4.1 *Curation of Ivuna stones BM1996 M4 and BM2008 M1*

All meteorites are susceptible to terrestrial alteration as soon as they enter the Earth's atmosphere (e.g. Bland et al. 2006). The CI carbonaceous chondrites are all falls (excluding the Antarctic CI finds that probably belong to separate chondrite group, King et al. 2019b), but as highly porous and volatile-element rich rocks they are particularly vulnerable to terrestrial modification and are known to experience mineralogical and chemical changes (Berzelius 1834; Zolensky 1999; Gounelle and Zolensky 2001; Velbel and Palmer 2011). Understanding the consequences and timescales of terrestrial alteration is important because CI chondrites are rare tracers of conditions and processes in the early protoplanetary disk, and current and future space missions are set to return similar materials from other solar system bodies.

The most obvious manifestation of terrestrial alteration in the CI chondrites is the presence of white sulphate efflorescence on meteorite surfaces or in-filling interior veins (Berzelius 1834; Zolensky 1999; Gounelle and Zolensky 2001; Velbel and Palmer 2011). Sulphate grains, which for Orgueil were reported only two weeks after its fall in 1864 (Daubrée 1864; Pisani 1864), and veins were originally attributed to fluid flow on the CI parent body (DuFresne and Anders 1961, 1962; McSween and Richardson 1977; Richardson 1978; Frederiksson and Kerridge 1988). However, Gounelle and Zolensky (2001) convincingly argued that a significant fraction of the sulphates in CI chondrites are the product of terrestrial alteration. Sulphates are highly reactive at room temperature and Gounelle and Zolensky (2001) showed sulphate formation on the surface of both a polished section of Orgueil (see their Fig. 2) and also a larger stone of the meteorite stored in a bell-jar, possibly since its fall, at the Musée d'Histoire Naturelle, Paris (see their Fig. 3). We see similar white efflorescence on the surface of the Ivuna BM2008 M1 stone (Fig. 1) and observed sulphate grains growing out of the Ivuna BM1996 M4 thin section (Fig. 5). The sulphate grains on BM1996 M4 formed within six years of the section being made, ~3 times faster than those reported on the section of Orgueil by Gounelle and Zolensky (2001).

Cracks and fractures in the polished section of Ivuna BM2008 M1 are enriched in Na and S, which are likely in Mg- and/or Na-bearing sulphates such as epsomite or bloedite. The BM2008 M1 section was analysed immediately after preparation and no sulphate grains were found on its surface, suggesting that the veins were present prior to the section being made. Sulphate veins are common in CI chondrites, with DuFresne and Anders (1961, 1962) providing the first reports of veins in Orgueil and Ivuna (97 and 24 years after their falls respectively). The formation of sulphate veins requires either the remobilization of elements within a meteorite or an addition of elements from the local terrestrial environment (Velbel and Palmer 2011). Richardson (1978) first showed that the matrix of CI chondrites is depleted in leachable elements including Na, S and Ca. Gounelle and Zolensky (2001) proposed that sulphates and sulphides formed on the parent body of the CI chondrites react with oxygen and atmospheric water, dissolve and become remobilised to fill the abundant pore spaces and cracks.

The presence of sulphate veins in BM2008 M1 but not BM1996 M4 perhaps reflects differences in the terrestrial history of these stones. Sulphate efflorescence is common on the surfaces of carbonaceous chondrites recovered from Antarctica that have terrestrial residences times of ~$10^5$ years (Bland et al. 2006). In contrast, Ivuna was recovered ~24 hours after falling to Earth (Oates 1941). The weather conditions at the time Ivuna was collected are unknown but Lake Rukwa in Tanzania has a tropical climate, with December part of the rainy season and having daily temperatures of ~15 – 25°C. In this environment a fragile meteorite such as Ivuna would have experienced some degree of terrestrial alteration despite its rapid recovery. Indeed, sulphates were quickly identified in Alais (Cuvier 1824; Berzelius 1834) and Orgueil (Daubrée 1864; Pisani 1864) after their fall, and it has been shown that only brief (days to years) exposure to the terrestrial environment is sufficient to initiate the process of weathering in chondritic meteorites (Gibson and Bogard 1978; Bland et al. 1998a; Barrat et al. 1999; Lee and Bland 2004; Velbel 2014; Walker et al. 2018). Nevertheless, the BM2008 M1 and BM1996 M4 stones would have undergone the same level of terrestrial modification prior to Ivuna being recovered.

Following its initial recovery there appears to be no description of how Ivuna was divided into multiple stones or how the pieces were later curated. The only curatorial records available for the Ivuna stones studied in this work indicate that BM1996 M4 has been stored either in air or wrapped in fluoropolymer film within wooden cabinets, and that BM2008 M1 has been stored in a nitrogen atmosphere within a sealed case at least since 2008 and potentially for longer. Subsequently we assume that during their terrestrial residence both the BM2008 M1 and BM1996 M4 stones probably experienced variable temperatures and humidity as their storage conditions changed. Such changes, even if for a short period of time, can continue the weathering process in a laboratory or museum environment. For example, Losiak and Velbel (2011) reported that Antarctic meteorites stored in a freezer that suffered a power failure had a higher-than-average abundance of evaporite minerals such as sulphates. Furthermore, the weathering rate of the BM2008 M1 and BM1996 M4 stones will also be a function of factors such as sulphate and sulphide abundances, porosity, and the mechanical properties of different clasts (e.g. Losiak and Velbel 2011), which can be highly variable in the heterogeneous CI chondrites (see section 4.2).

Once terrestrial alteration of a CI chondrite has begun, limiting the effects may be challenging. Recrystallization of sulphates into pore spaces, cracks and veins induces stresses that can open-up new fractures and facilitates further weathering (Gounelle and Zolensky 2001). Storing CI chondrites in a nitrogen atmosphere may help to suppress this process, although we still observe white efflorescence on the surface and sulphate veins within the interior of the Ivuna stone BM2008 M1. While it is likely that the alteration of the BM2008 M1 stone had started prior its arrival at the NHM, Jull et al. (1988) showed that white efflorescent weathering products can form on ordinary chondrites stored under dry nitrogen in as little as nine months. This issue is potentially even worse for the highly porous and volatile-element rich CI chondrites and implies that the only way to maintain the pristine nature of samples from hydrated carbonaceous asteroids is to minimise their exposure to the terrestrial environment and carefully curate them in a stable, preferably cold ($\sim -10$°C) environment (e.g. Herd et al. 2016).

### 4.2 *Brecciation and Sample Heterogeneity*

The CI chondrites are known to be extremely heterogeneous having undergone extensive aqueous alteration and impact brecciation (Tomeoka and Buseck 1988; Endress and Bischoff 1996; Morlok et al. 2006; Barrat et al. 2012). Morlok et al. (2006) reported numerous fragments and clasts in the CI chondrites ranging in size from 40 µm up to several hundred µm with a wide range of mineralogical and chemical compositions. In total they identified at least eight distinct lithologies (e.g. Mg-rich phyllosilicate, Fe-rich phyllosilicate, phosphate-rich etc.) and argued that the heterogeneity of the CI chondrites reflected aqueous alteration of different starting materials followed by mixing together of clasts by impacts. Our observations of the Ivuna stones BM1996 M4 and BM2008 M1 are consistent with them being breccias - both samples contain mm-sized clasts with variable phyllosilicate compositions and abundances of carbonates, sulphides and magnetite.

We note that the individual clasts both in our study (e.g. see Fig. 3) and described by Morlok et al. (2006) are often rounded. This rounding could result from a mechanical process induced by impacts, with the CI chondrites interpreted as regolith breccias due to the presence of a trapped solar wind component (Bischoff et al. 2006). Alternatively, we speculate that the rounded shapes of the clasts, plus a lack of any obvious shock features, instead suggest that these could be samples from the parent body interior. In this case the brecciation is formed by gentle convection driven by radioactive heating of icy primordial clasts, in a similar process to that proposed by Bland and Travis (2017).

The heterogeneous nature of the CI chondrites and irregular distribution of minerals such as carbonates and phosphates has important implications for interpreting bulk chemical datasets. The CI chondrites are clearly heterogeneous at the mm-scale and therefore sub-samples for chemical analyses should ideally be larger than this in order to be representative of the bulk meteorite. For a CI chondrite, a mass of ~2 mg (1000 $\mu m^3$) potentially samples only a single clast, and even larger samples may be subject to homogeneity issues if they contain single grains with unusual chemical or isotopic characteristics. This “nugget effect” has been observed and discussed in other aqueously altered carbonaceous chondrites (e.g. Dauphas and Pourmand 2015; King et al. 2019a).

Many chemical studies of the CI chondrites analysed samples with masses <200 mg, leading to results that deviated significantly from CI-compositions (see Section 3.3; Ebihara et al. 1982; Rocholl and Jochum 1993; Pourmand et al. 2012). Morlok et al. (2006) calculated that ~1–2 g is the minimum mass of material needed for representative chemical analyses of the CI chondrites. Barrat et al. (2012) analysed multiple ~0.5–1 g samples of Orgueil, Ivuna and Alais and still found evidence for chemical heterogeneity in elements such as Na, K, Rb, Cs and U that could be related to local differences in the degree of aqueous alteration (Blinova et al. 2014). However, in view of the scarcity of CI material, it is not always possible to obtain a sample of sufficient mass to be an adequate representation of the bulk, and in this study we were limited to ~40 mg aliquots of Ivuna samples for each type of analysis, which is clearly below the mass threshold suggested by Morlok et al. (2006) and Barrat et al. (2012). Even though the chemical composition of each sample will be strongly influenced by its mineral contents, the validity of using such small aliquots for chemical analysis is supported by King et al. (2015b), who found only minor variations (<4 vol%) in the mineralogy of multiple ~50 mg aliquots of Ivuna and Orgueil.

There are several discrepancies between the Ivuna stones BM1996 M4 and BM2008 M1, and between our samples and previous analyses of Ivuna. The most striking is the large enrichment of the heavy REEs (Dy – Lu) and Zr, Hf, Ta, Th, and U in BM1996 M4 (Fig. 10). This suggests either sample contamination (which is unlikely) or a mineralogical control. Morlok et al. (2006) reported that some phosphate grains in Orgueil are significantly enriched in heavy REEs with Lu/La ratios up to 1.9. Because the REE concentrations in these phosphate grains are as much as 150 times the average CI level, their presence (undetected by, for example, bulk P content) could easily explain the observed Lu/La ratio of 1.6 in BM1996 M4.

Similar enrichment of REEs and Zr, Hf Ta, Th and U was reported in samples of the aqueously altered CM chondrite Jbilet Winselwan and attributed to ultrarefractory inclusions (Friend et al. 2018; King et al. 2019a). However, CAIs are generally considered to be absent from the CI chondrites (MacPherson et al. 2005), although some refractory oxide grains have been found in Orgueil (Huss et al. 1995). This, along with our observed heterogeneities in trace refractory elements, suggests that CIs may have once contained CAIs that were entirely destroyed by pervasive aqueous alteration. Regardless of the cause, the heavy REEs and Zr, Hf, Ta, Th, and U abundances in BM1996 M4 are all outliers according to Grubbs' test and were excluded from calculating the average Ivuna chemical composition (see Section 3.3).

### 4.3 *Comparison of new average Ivuna composition with published abundances of other CI chondrites and the Sun*

Figure 11 compares our new averaged Ivuna mass concentrations to those given by Lodders (2003) for specifically the Ivuna meteorite, which were evaluated from an average of published values that were available at that time. There is a close correspondence between the datasets, with only Ba and Tb having deviations of more than 20%. However, the Lodders (2003) Ivuna values for these elements are based on single measurements only.

The average Ivuna composition derived in Section 3.3 was converted to element abundances for comparison with meteoritic and solar photospheric abundances. As discussed by Lodders (2003), because hydrogen is depleted in meteorites, meteoritic abundances are commonly expressed as mass fractions (ppm or μg $g^{-1}$) or, for comparison purposes, a cosmochemical scale (numbers of atoms with Si = $10^6$ atoms). An astronomical abundance scale, which is on a logarithmic number of atoms scale with H = 12, additionally allows comparison with the extensive data sets available of measured photospheric abundances (e.g., Asplund et al. 2009). To convert meteoritic cosmochemical abundances to an astronomical scale, we used the photospheric abundance of Si recommended by Scott et al. (2015a), $A(\mathrm{Si}) = 7.51 \pm 0.03$, to obtain:

$$A(\mathrm{El}) = 1.51 + N(\mathrm{El})$$

where $N$ is the CI chondrite abundance on a cosmochemical scale (atom number density of an element El relative to Si with $10^6$ atoms). This is in agreement with Asplund et al. (2009) but is slightly different from Lodders (2003) because of the use of an earlier photospheric Si abundance. Such a normalization with the photospheric Si abundance should show any abundance differences for individual elements. The cosmochemical and astronomical abundances for the new averaged Ivuna concentrations are given in Table 2. Also given (on an astronomical scale) in Table 2 are the most recent estimates of solar photospheric abundances, and the meteoritic abundances from measurements based on the larger and much more widely studied Orgueil CI chondrite (Lodders et al. 2009).

Figure 12 compares our new average Ivuna abundances with those for Orgueil given by Lodders et al. (2009) for 53 elements. The vertical error bars on Fig. 12 are the combined standard deviations (σ) taken to be the standard deviations of the Ivuna averaged abundances ($\sigma_I$) and those of Lodders et al. (2009) ($\sigma_L$) summed quadratically, $\sigma = (\sigma_I^2 + \sigma_L^2)^{1/2}$. There are differences of between 0.05 and 0.15 in the logarithm for five elements, Be, P, K, Ba and W; these are larger than the combined standard deviations for Be, Ba and W. The largest deviation (41%, or 0.15 in the logarithm) is for Ba, which is more abundant in the Ivuna average. The combined standard deviation in the Ivuna abundances is comparatively large mostly because of the 24% spread of Ba values in forming the Ivuna average, although there are no outliers according to Grubbs' test. Elevated Ba abundances are a common feature of hot desert weathering (e.g. Crozaz et al. 2003), but Ivuna is a fall and other elements such as Sr and U show no significant enrichment. A more likely explanation is that different stones of Ivuna contain variable amounts of carbonate and phosphate grains.

Figure 12 shows evidence of a slight trend in the data for the heavy REEs (Ho – Lu, $Z$ = 67 – 71) but it is less than the combined standard deviations. We note that there is agreement between the Ivuna average and the Lodders et al. (2009) abundances for the REEs Sc ($Z$ = 21) and Y ($Z$ = 39) despite the exclusion of some measurements from the Ivuna average. In summary, we find that our new Ivuna average composition is not significantly different from the compilation by Lodders et al. (2009) based on the Orgueil meteorite.

In general, photospheric abundances are estimated from measurements of solar absorption lines together with atomic data (in the form of transition probabilities) and standard solar atmospheres. A steady improvement in recent years in both atomic data and atmospheric models including most notably the introduction of three-dimensional hydrodynamic codes has led to significant revisions, even for high-abundance elements such as C, N, and O. The most recent works on photospheric abundances, updating the work of Asplund et al. (2009), include those by Grevesse et al. (2015: Cu – Th, $Z$ = 29 – 90), Scott et al. (2015a: Na – Ca, $Z$ = 11 – 20), and Scott et al. (2015b: Sc – Ni, $Z$ = 21 – 28). Photospheric abundance determinations are not possible for Sb ($Z$ = 51), Cs ($Z$ = 55), and Ta ($Z$ = 73) owing to a lack of absorption lines in photospheric spectra (Asplund et al. 2009).

Figure 13 compares the Ivuna averaged abundances with photospheric abundances for 49 elements, with error bars equal to the standard deviations in the Ivuna averages and photospheric estimates combined quadratically. The main difference (-2.22 ± 0.11) is Li, the CI chondrite abundance estimated here being a factor of 166 (with limits defined by the standard deviations in the mass fractions of from 128 to 214) more than the photospheric abundance. The Lodders et al. (2009) Li abundance for Orgueil is very close to that derived here. The depletion in the solar photospheric Li abundance is well documented although its exact cause remains under discussion. It may be due to mixing just below the solar convective zone (where the temperature is approximately 2 MK) to a level where the temperature is 2.5 MK (Charbonnel and Talon 2005) and Li is destroyed. Alternatively, as in all planet-bearing stars, a deepening of the convective zone to a temperature of ~2.5 MK may occur by the presence of planets (Israelian et al. 2009). Our Be abundance (1.37 ± 0.01) is higher than that of Lodders et al. (2009) (1.32 ± 0.03), but close to the value of Asplund et al. (2009) (1.38 ± 0.09). Thus, although photospheric Li is depleted compared with CI chondrites, Be, which is destroyed in the solar interior at a temperature of ~3.5 MK, apparently is not. This result is consistent with the work of Balachandran and Bell (1998) in which the photospheric Be abundance, based on ionized Be lines in the near-ultraviolet, was revised upwards to meteoritic values to account for opacity effects, and suggests that any turbulence below the solar convection zone occurs between 2.5 MK and 3.5 MK.

Other differences (of ~26% or 0.1 in the logarithm) between the Ivuna average and photospheric abundances occur for Sc, Rb, Mo, Ag, Tl, and Pb ($Z$ = 21, 37, 47, 81, 82), the largest being for Sc, Ag and Pb. For Sc, the difference between photospheric and meteoritic abundances has been remarked on by Asplund et al. (2009) and Scott et al. (2015b). The difference is larger than the combined uncertainties, and is confirmed here, although there is no apparent explanation. For Ag, the photospheric abundance is less than the Ivuna average abundance by 0.23 (±0.12) in the logarithm. Our value is confirmed by the Lodders et al. (2009) Orgueil value. Possibly the uncertainty in the photospheric abundance, which is based on two neutral Ag lines in the near-ultraviolet that are difficult to measure (Grevesse et al. 2015), is the reason for the difference. For Pb, the photospheric abundance relies on a single line in a crowded spectral region, as discussed by Grevesse et al. (2015), so it appears likely that this explains the differences.

In summary, we conclude that the most recent photospheric abundances for the 49 elements for which the comparison was possible differ from the average Ivuna bulk composition by amounts that are less than the combined uncertainties with the possible exception of Ag and Pb for which the photospheric abundance is poorly constrained.

## 5. Conclusions

Modification by the terrestrial environment significantly hampers our ability to decipher the geological record of carbonaceous chondrite meteorites, and is also a major challenge for future analysis of extra-terrestrial materials returned from other solar system bodies. The rare CI carbonaceous chondrites are both the most chemically primitive and aqueously altered meteorite group but despite them all being falls, these fragile, porous and volatile-element rich rocks are highly susceptible to terrestrial weathering processes. The CI type specimen, Ivuna, is the most recent fall of appreciable size (~0.7 kg) and therefore likely to be the CI chondrite least affected by the terrestrial environment and available for the community to study. Here, we present the results of a detailed petrographic investigation of two Ivuna stones – BM1996 M4 and BM2008 M1 – currently curated at the NHM, London. The BM1996 M4 stone has been stored either in air or wrapped in Teflon within wooden cabinets, while BM2008 M1 has been kept in a nitrogen atmosphere at least since its arrival at the NHM (possibly longer), making it potentially the most pristine piece of a CI chondrite. In addition, due to the relative scarcity of chemical data in the literature for Ivuna we also present bulk major, minor and trace element abundances for both BM1996 M4 and BM2008 M1. Our main conclusions are:-

1. In thin section, the Ivuna stones BM1996 M4 and BM2008 M1 are breccias with multiple rounded clasts containing abundant fine-grained phyllosilicates in which are embedded varying abundances of carbonates, sulphides and magnetite. These observations are consistent with previously published literature on Ivuna and the other CI chondrites. We propose that the rounding of the clasts may have been induced by impact gardening of the parent body surface or through convection driven by radioactive heating of icy primordial clasts.

2. Sulphate grains are observed growing out of the polished section of BM1996 M4. The grains must have formed in ≤6 years, indicating sulphate formation on the surface of thin sections ~3 times faster than previous reports for CI chondrites. In comparison, the sulphates in BM2008 M1 in-fill cracks and fractures, a feature that was not identified in BM1996 M4. The sulphate veins in BM2008 M1 formed when primary sulphates and sulphides in the sample reacted with oxygen and atmospheric water, leading to dissolution and remobilisation of secondary sulphates into the abundant pores and fractures.

3. Differences in the extent of terrestrial alteration recorded by BM1996 M4 and BM2008 M1 probably reflect variations in mineralogy, porosity, and perhaps most importantly the post-recovery curation history, which until arrival at the NHM was poorly documented for these Ivuna stones. While the effects of nitrogen storage on suppressing the terrestrial alteration of carbonaceous chondrites need to be explored further, currently the only way to ensure the pristine nature of samples from hydrated carbonaceous asteroids is to avoid exposure to the terrestrial environment in a stable environment.

4. In general we find good agreement between the major, minor, and trace element concentrations in BM1996 M4 and BM2008 M1 and previously published bulk chemical analyses of the Ivuna meteorite. The only significant discrepancy is a large enrichment of the heavy REEs (Dy – Lu) and Zr, Hf, Ta and U in BM1996 M4, which might result from variations in the distribution of phosphate grains in the sample. This reinforces the idea that CI chondrites are heterogeneous at the mm-scale and that where possible samples for chemical analyses should ideally be larger than this to be representative of the bulk meteorite.

5. Chemical analyses of BM1996 M4 and BM2008 M1 combined with published analytical data to calculate a new average composition for the Ivuna meteorite show good agreement with previous compilations of element abundances in Ivuna and Orgueil except for Ba, which likely reflects variable amounts of carbonate and/or phosphate grains between the different analysed samples.

6. Comparison between our average Ivuna composition and the most recent solar photospheric abundances shows insignificant differences except for Ag and Pb. These discrepancies are probably due to uncertainties in determining the photospheric abundances of these elements. Lithium is clearly depleted in the photosphere relative to its CI abundance whereas Be abundances are similar in CI chondrites and the photosphere, suggesting that any turbulence below the solar convection zone occurs between 2.5 MK and 3.5 MK.

## Acknowledgements

We thank Yann Le Gac for contributing SEM and EDS images to the project. Caroline Smith and Natasha Almeida are thanked for their role in acquiring the Ivuna sample BM2008 M1 on behalf of the Natural History Museum, London, and for their curatorial role. Emma Humphries-Williams is thanked for assistance with the ICP-OES analysis. The manuscript was significantly improved by the comments of Michael Velbel, Chris Herd and Eric Quirico. This work was supported by the Science and Technology Facilities Council (STFC), UK, through grants ST/J001473/1 and ST/M00094X/1.

**Figure 1.** The main mass of the Ivuna CI carbonaceous chondrite at the Natural History Museum, London. The sample (BM2008 M1) is ~5 cm in maximum dimension, weighs ~200 g, and is stored in an inert nitrogen atmosphere within a sealed case. The white efflorescence that can be seen on the meteorite surface is probably a sulphate phase.

**Figure 2.** BSE image of a thin section of Ivuna BM1996 M4 (P12628). The image was collected using a Zeiss EVO 15LS SEM six years after the section was made. Note evidence for brecciation, e.g. patches where bright sulphides are more abundant.

**Figure 3**. X-ray element map of Ivuna BM1996 M4 showing the distribution of Si (blue), S (green), Ca (yellow) and K (red). In the map sulphide laths show up as bright green and carbonates as yellow grains disseminated throughout the field of view. Clasts (often rounded in shape) can be easily distinguished by variations in the abundance of K-bearing phases (pink). The map was acquired using an FEI Quanta 650 FEG SEM at 6 kV.

**Figure 4.** BSE image of tufty, fibrous, Fe-rich phyllosilicates in Ivuna BM1996 M4. The image was acquired using an FEI Quanta 650 FEG SEM at 9 kV.

**Figure 5.** SE image (a) and X-ray element map (b) of a sulphate grain that has grown on a phyllosilicate substrate in Ivuna BM1996 M4. The image and map were acquired using an FEI Quanta 650 FEG SEM at 6 kV.

**Figure 6**. X-ray map showing the distribution of carbon in Ivuna BM1996 M4 (same field of view as Fig. 3). Hotspots of carbon (red) are mainly concentrated along cracks and fractures but also disseminated throughout the field of view. The map was acquired using an FEI Quanta 650 FEG SEM at 6 kV.

**Figure 7.** BSE image of the entire Ivuna BM2008 M1 thin section (P16384). The image was collected using a Zeiss EVO 15LS SEM immediately after preparation to avoid degradation of the sample. Note evidence for brecciation, e.g. patches where bright sulphides are more abundant.

**Figure 8**. X-ray element map showing the distribution of Mg (red), Al (blue) and Ca (green) overlain on a BSE image of Ivuna BM2008 M1. The dark red background is phyllosilicate-

rich, sulphides are bright white, and the green and light blue patches are probably carbonate grains. Sulphides and carbonates are finely disseminated throughout the thin section. The image and element map were collected using a Zeiss EVO 15LS SEM immediately after preparation to avoid degradation of the sample.

**Figure 9.** BSE image (a) and X-ray element maps showing the distribution of S (b), Na (c) and Ca (d) in Ivuna BM2008 M1. The BSE image is fairly featureless except for sulphide grains scattered throughout the matrix and cracks and fractures that are enriched in S and Na but typically not Ca. The image and element maps were collected using a Zeiss EVO 15LS SEM immediately after preparation to avoid degradation of the sample.

**Figure 10**. Comparison between the major (a), REE (b) and trace element (c) abundances measured in the Ivuna stones BM1996 M4 and BM2008 M1 as part of this study, and those reported for Ivuna by Barrat et al. (2012).

**Figure 11**. Mass fractions (ppm or $\mu g\ g^{-1}$) given for the Ivuna meteorite by Lodders et al. (2003: Table 3) plotted logarithmically against the Ivuna averaged values in Table 1. The uncertainties are smaller than the size of the symbols.

**Figure 12.** Abundances for 53 elements (logarithmic scale, H = 12) based on the compilation for the Orgueil CI chondrite made by Lodders et al. (2009: Table 4) minus those derived here for the Ivuna average (Table 1) plotted against atomic number (Z). Error bars are the uncertainties in the Lodders et al. (2009) compilation and those in Table 1 combined quadratically.

**Figure 13.** Abundances (logarithmic scale, H = 12) based on the Ivuna average derived here minus photospheric values (see Table 2) for 49 elements plotted against atomic number (Z). Apart from Li (off-scale, photospheric abundance a factor of 166 less than Ivuna), the main differences are for Ag (Z = 47) and Pb (Z = 82).

**Table 1**. Element abundances in the Ivuna meteorite.

| Weight | LOQ (%) | No. of analyses | Ivuna BM1996 (average) | S.D. | Ivuna BM2008 (average) | S.D. | Barrat et al. (2012) | No. of analyses used to calculate average | Ivuna average | S.D. | R.S.D. |
|---|---|---|---|---|---|---|---|---|---|---|---|
| ***Major and Minor Elements*** | | | | | | | | | | | |
| $Al_2O_3$ | 0.05 | 3 | 1.708 | 0.013 | 1.683 | 0.019 | 1.48 | 8 | **1.596** | 0.081 | 5% |
| CaO | 0.05 | 3 | 1.238 | 0.008 | 1.314 | 0.011 | 1.29 | 5 | **1.290** | 0.048 | 4% |
| $Fe_2O_3$ | 0.01 | 3 | 24.3 | 0.12 | 24.4 | 0.22 | 27.0 | 8 | **25.2** | 1.4 | 5% |
| MgO | 0.01 | 3 | 14.58 | 0.12 | 14.62 | 0.14 | 15.84 | 6 | **15.28** | 0.86 | 6% |
| $Na_2O$ | 0.05 | 3 | 0.584 | 0.004 | 0.843 | 0.010 | 0.661 | 11 | **0.706** | 0.081 | 12% |
| $P_2O_5$ | 0.1 | 3 | 0.1866 | 0.0007 | 0.164 | 0.0019 | 0.234 | 5 | **0.190** | 0.027 | 14% |
| $SiO_2$ | 0.1 | 3 | 21.5 | 0.11 | 21.15 | 0.12 | | 5 | **22.07** | 0.89 | 4% |
| $TiO_2$ | 0.01 | 3 | 0.0664 | 0.0005 | 0.0675 | 0.0036 | 0.072 | 6 | **0.0678** | 0.0032 | 5% |
| ***REEs*** | ppm | | | | | | | | | | |
| La | 0.0020 | 3 | 0.238 | 0.003 | 0.234 | 0.003 | 0.241 | 8 | **0.242** | 0.027 | 11% |
| Ce | 0.0400 | 3 | 0.700 | 0.008 | 0.645 | 0.007 | 0.614 | 9 | **0.635** | 0.060 | 9% |
| Pr | 0.0006 | 3 | 0.0989 | 0.0017 | 0.0896 | 0.0026 | 0.092 | 7 | **0.0963** | 0.0086 | 9% |
| Nd | 0.0040 | 3 | 0.505 | 0.015 | 0.464 | 0.003 | 0.471 | 9 | **0.480** | 0.043 | 9% |
| Sm | 0.0026 | 3 | 0.159 | 0.002 | 0.145 | 0.001 | 0.152 | 8 | **0.156** | 0.010 | 7% |
| Eu | 0.0008 | 3 | 0.0638 | 0.0024 | 0.0582 | 0.0037 | 0.0597 | 9 | **0.0591** | 0.0055 | 9% |
| Gd | 0.0007 | 3 | 0.224 | 0.005 | 0.195 | 0.004 | 0.213 | 8 | **0.212** | 0.012 | 6% |
| Tb | 0.0002 | 3 | 0.0413 | 0.0011 | 0.0352 | 0.0007 | 0.0388 | 8 | **0.0376** | 0.0034 | 9% |
| Dy | 0.0017 | 3 | 0.296 | 0.004 | 0.237 | 0.004 | 0.26 | 8 | **0.259** | 0.021 | 8% |
| Ho | 0.0004 | 3 | 0.0713 | 0.0017 | 0.0541 | 0.0011 | 0.0579 | 6 | **0.0585** | 0.0071 | 12% |
| Er | 0.0009 | 3 | 0.216 | 0.002 | 0.158 | 0.003 | 0.165 | 8 | **0.163** | 0.012 | 7% |
| Tm | 0.0002 | 3 | 0.0358 | 0.0017 | 0.0238 | 0.0004 | 0.0271 | 7 | **0.0256** | 0.0024 | 9% |
| Yb | 0.0015 | 3 | 0.246 | 0.001 | 0.159 | 0.002 | 0.171 | 9 | **0.166** | 0.013 | 8% |
| Lu | 0.0009 | 3 | 0.0396 | 0.0012 | 0.0243 | 0.0007 | 0.0248 | 8 | **0.0250** | 0.0031 | 12% |

***Trace Elements***

| | | | | | | | | | | | |
|---|---|---|---|---|---|---|---|---|---|---|---|
| Li | 0.0688 | 1 | 1.44 | | 1.47 | | 1.44 | 5 | **1.47** | 0.04 | 3% |
| Be | 0.0050 | 3 | 0.0243 | 0.0014 | 0.0238 | 0.0015 | 0.0232 | 3 | **0.0238** | 0.0006 | 2% |
| P | | 3 | 851 | 3 | 746 | 8 | 1020 | 5 | **844** | 109 | 13% |
| K | 39 | 1 | 330 | | 521 | | 432 | 8 | **472** | 91 | 19% |
| Sc | 0.0352 | 3 | 5.61 | 0.05 | 5.60 | 0.01 | 6.08 | 7 | **5.62** | 0.48 | 9% |
| Ti | 2.0 | 1 | 465 | | 460 | | 466 | 6 | **446** | 37 | 8% |
| V | 0.0410 | 1 | 48.8 | | 48.7 | | 54 | 6 | **51.4** | 4.5 | 9% |
| Cr | 2.0 | 1 | 2303 | | 2329 | | 2570 | 8 | **2427** | 143 | 6% |
| Mn | 1.0 | 1 | 1855 | | 1772 | | 2003 | 8 | **1868** | 113 | 6% |
| Co | 0.1003 | 1 | 525 | | 513 | | 554 | 8 | **492** | 42 | 9% |
| Ni | 1.1 | 1 | 11448 | | 10765 | | 12000 | 6 | **10999** | 606 | 6% |
| Cu | 0.1781 | 1 | 131 | | 125 | | 138 | 6 | **132.3** | 5.4 | 4% |
| Zn | 0.0610 | 1 | 307 | | 306 | | 330 | 10 | **306** | 19 | 6% |
| Ga | 0.0060 | 3 | 9.29 | 0.05 | 9.16 | 0.03 | 9.67 | 5 | **9.12** | 0.39 | 4% |
| Rb | 0.0109 | 4 | 2.17 | 0.08 | 2.22 | 0.13 | 2.23 | 10 | **2.18** | 0.29 | 13% |
| Sr | 0.0485 | 1 | 7.61 | | 7.40 | | 8.04 | 4 | **7.75** | 0.30 | 4% |
| Y | 0.0052 | 3 | 1.52 | 0.02 | 1.136 | 0.003 | 1.6 | 4 | **1.571** | 0.062 | 4% |
| Zr | 0.20 | 3 | 13.9 | 0.08 | 3.92 | 0.007 | 3.48 | 4 | **3.57** | 0.27 | 8% |
| Nb | 0.0087 | 3 | 0.273 | 0.003 | 0.272 | 0.001 | 0.298 | 6 | **0.274** | 0.022 | 8% |
| Mo | 0.0215 | 3 | 0.861 | 0.001 | 0.897 | 0.009 | | 4 | **1.11** | 0.30 | 27% |
| Ag | 0.0131 | 1 | 0.205 | | 0.223 | | | 6 | **0.195** | 0.028 | 14% |
| Cd | 0.0100 | 1 | 0.790 | | 0.740 | | | 6 | **0.701** | 0.055 | 8% |
| Sn | 0.30 | 3 | 1.77 | 0.03 | 1.94 | 0.04 | | 6 | **1.73** | 0.16 | 9% |
| Sb | 0.0331 | 3 | 0.134 | 0.003 | 0.129 | 0.002 | | 10 | **0.160** | 0.031 | 19% |
| Cs | 0.0092 | 3 | 0.205 | 0.022 | 0.202 | 0.0110 | 0.192 | 9 | **0.193** | 0.017 | 9% |
| Ba | 0.0256 | 3 | 3.90 | 0.07 | 4.38 | 0.06 | 2.57 | 5 | **3.21** | 0.91 | 28% |
| Hf | 0.0341 | 3 | 0.413 | 0.004 | 0.144 | 0.002 | 0.107 | 4 | **0.112** | 0.022 | 20% |
| Ta | 0.0120 | 3 | 0.0380 | 0.0010 | 0.0155 | 0.0004 | 0.0149 | 3 | **0.0149** | 0.0006 | 4% |
| W | 0.0378 | 3 | 0.131 | 0.002 | 0.136 | 0.004 | 0.1 | 4 | **0.122** | 0.019 | 16% |
| Tl | 0.0020 | 1 | 0.165 | | 0.138 | | | 5 | **0.142** | 0.015 | 10% |
| Pb | 0.0184 | 1 | 3.11 | | 3.11 | | 2.65 | 5 | **2.75** | 0.34 | 12% |
| Th | 0.0132 | 3 | 0.0613 | 0.0014 | 0.0346 | 0.0003 | 0.0289 | 7 | **0.0290** | 0.0029 | 10% |
| U | 0.0009 | 3 | 0.0198 | 0.0014 | 0.01066 | 0.0005 | 0.00749 | 9 | **0.0085** | 0.0021 | 24% |

Table gives major, minor and trace element abundances (ppm or µg $g^{-1}$) measured in the Ivuna stones BM1996 M4 and BM2008 M1, and our new averaged Ivuna abundances. The averaged values are based on the present work and published data of Barrat et al. (2012), Greenland (1967), Wolf and Palme (2001), Wiik (1956), Case et al. (1973), Krähenbühl et al. (1973), Ebihara et al. (1982), Rochol and Jochum (1993), Schmitt et al. (1964), Pourmand et al. (2012), Kaushal and Wetherill (1970) and Braukmüller et al. (2018). LOQ = limit of quantification; S.D. = standard deviations; R.S.D. = relative standard deviations.

**Table 2.** Ivuna element abundances compared with solar system values.

| Z | Element | Cosmochemical | S.D. | Astronomical | S.D. | Photospheric | S.D. | Meteoritic | S.D. |
|---|---|---|---|---|---|---|---|---|---|
| 3 | Li | 57.7 | 1.6 | 3.27 | 0.01 | 1.05 | 0.1 | 3.28 | 0.05 |
| 4 | Be | 0.719 | 0.018 | 1.37 | 0.01 | 1.38 | 0.09 | 1.32 | 0.03 |
| 11 | Na | 62500 | 7500 | 6.31 | 0.05 | 6.21 | 0.04 | 6.29 | 0.02 |
| 12 | Mg | 1050000 | 30000 | 7.53 | 0.01 | 7.59 | 0.04 | 7.55 | 0.01 |
| 13 | Al | 86000 | 2300 | 6.44 | 0.01 | 6.43 | 0.04 | 6.45 | 0.01 |
| 14 | Si | 1000000 | 18800 | 7.51 | 0.01 | 7.51 | 0.03 | 7.53 | 0.01 |
| 15 | P | 7420 | 1110 | 5.38 | 0.06 | 5.41 | 0.03 | 5.45 | 0.04 |
| 19 | K | 3360 | 660 | 5.04 | 0.08 | 5.04 | 0.05 | 5.1 | 0.02 |
| 20 | Ca | 63100 | 170 | 6.31 | 0.01 | 6.32 | 0.03 | 6.31 | 0.02 |
| 21 | Sc | 34 | 2.9 | 3.04 | 0.04 | 3.16 | 0.04 | 3.07 | 0.02 |
| 22 | Ti | 2630 | 20 | 4.93 | 0 | 4.93 | 0.04 | 4.93 | 0.03 |
| 23 | V | 278 | 21 | 3.95 | 0.03 | 3.89 | 0.08 | 3.98 | 0.02 |
| 24 | Cr | 12600 | 740 | 5.61 | 0.02 | 5.62 | 0.04 | 5.66 | 0.01 |
| 25 | Mn | 9130 | 480 | 5.47 | 0.02 | 5.42 | 0.04 | 5.5 | 0.01 |
| 26 | Fe | 865000 | 36000 | 7.44 | 0.02 | 7.47 | 0.04 | 7.47 | 0.01 |
| 27 | Co | 2260 | 210 | 4.86 | 0.04 | 4.93 | 0.05 | 4.89 | 0.01 |
| 28 | Ni | 51000 | 2800 | 6.22 | 0.02 | 6.2 | 0.04 | 6.22 | 0.01 |
| 29 | Cu | 566 | 26 | 4.26 | 0.02 | 4.18 | 0.05 | 4.27 | 0.04 |
| 30 | Zn | 1270 | 80 | 4.61 | 0.03 | 4.56 | 0.05 | 4.65 | 0.04 |
| 31 | Ga | 36 | 1.4 | 3.07 | 0.02 | 3.02 | 0.05 | 3.1 | 0.02 |
| 37 | Rb | 6.98 | 0.96 | 2.35 | 0.06 | 2.47 | 0.07 | 2.38 | 0.03 |
| 38 | Sr | 24.1 | 0.9 | 2.89 | 0.02 | 2.83 | 0.06 | 2.9 | 0.03 |
| 39 | Y | 4.81 | 0.18 | 2.19 | 0.02 | 2.21 | 0.05 | 2.19 | 0.04 |
| 40 | Zr | 11.0 | 0.7 | 2.55 | 0.03 | 2.59 | 0.04 | 2.55 | 0.04 |
| 41 | Nb | 0.815 | 0.062 | 1.42 | 0.03 | 1.47 | 0.06 | 1.43 | 0.04 |
| 42 | Mo | 3.15 | 0.85 | 2.01 | 0.1 | 1.88 | 0.09 | 1.96 | 0.04 |
| 47 | Ag | 0.480 | 0.076 | 1.19 | 0.06 | 0.96 | 0.1 | 1.22 | 0.02 |
| 48 | Cd | 1.70 | 0.15 | 1.74 | 0.04 | 1.77 | 0.15 | 1.73 | 0.03 |
| 50 | Sn | 4.08 | 0.28 | 2.12 | 0.03 | 2.02 | 0.1 | 2.09 | 0.06 |
| 51 | Sb | 0.358 | 0.069 | 1.06 | 0.08 | -10 | 0 | 1.03 | 0.06 |
| 55 | Cs | 0.401 | 0.033 | 1.11 | 0.03 | -10 | 0 | 1.1 | 0.02 |
| 56 | Ba | 6.86 | 1.63 | 2.35 | 0.09 | 2.25 | 0.07 | 2.2 | 0.03 |
| 57 | La | 0.49 | 0.035 | 1.2 | 0.03 | 1.11 | 0.04 | 1.19 | 0.02 |
| 58 | Ce | 1.26 | 0.07 | 1.61 | 0.02 | 1.58 | 0.04 | 1.6 | 0.02 |
| 59 | Pr | 0.191 | 0.014 | 0.79 | 0.03 | 0.72 | 0.04 | 0.78 | 0.03 |
| 60 | Nd | 0.919 | 0.077 | 1.47 | 0.04 | 1.42 | 0.04 | 1.47 | 0.02 |
| 62 | Sm | 0.286 | 0.018 | 0.97 | 0.03 | 0.95 | 0.04 | 0.96 | 0.02 |
| 63 | Eu | 0.107 | 0.011 | 0.54 | 0.04 | 0.52 | 0.04 | 0.53 | 0.02 |
| 64 | Gd | 0.367 | 0.023 | 1.07 | 0.03 | 1.08 | 0.04 | 1.07 | 0.02 |
| 65 | Tb | 0.0651 | 0.00065 | 0.32 | 0.04 | 0.31 | 0.1 | 0.34 | 0.03 |
| 66 | Dy | 0.437 | 0.036 | 1.15 | 0.04 | 1.1 | 0.04 | 1.15 | 0.02 |
| 67 | Ho | 0.0974 | 0.0116 | 0.5 | 0.05 | 0.48 | 0.11 | 0.49 | 0.03 |
| 68 | Er | 0.267 | 0.021 | 0.93 | 0.03 | 0.93 | 0.05 | 0.94 | 0.02 |
| 69 | Tm | 0.0419 | 0.0048 | 0.13 | 0.27 | 0.11 | 0.04 | 0.14 | 0.03 |
| 70 | Yb | 0.261 | 0.024 | 0.93 | 0.04 | 0.85 | 0.11 | 0.94 | 0.02 |
| 71 | Lu | 0.0389 | 0.0047 | 0.1 | 0.05 | 0.1 | 0.09 | 0.11 | 0.02 |
| 72 | Hf | 0.178 | 0.037 | 0.76 | 0.08 | 0.85 | 0.05 | 0.73 | 0.02 |
| 73 | Ta | 0.0224 | 0.0009 | -0.14 | 0.02 | -10 | 0 | -0.14 | 0.04 |
| 74 | W | 0.178 | 0.030 | 0.76 | 0.07 | 0.83 | 0.11 | 0.67 | 0.04 |
| 81 | Tl | 0.192 | 0.021 | 0.79 | 0.05 | 0.9 | 0.2 | 0.79 | 0.03 |
| 82 | Pb | 3.73 | 0.42 | 2.08 | 0.05 | 1.92 | 0.08 | 2.06 | 0.03 |
| 90 | Th | 0.0352 | 0.0035 | 0.06 | 0.04 | 0.03 | 0.1 | 0.08 | 0.03 |
| 92 | U | 0.0101 | 0.0023 | -0.49 | 0.09 | -10 | 0 | -0.52 | 0.03 |

Table gives our new averaged Ivuna abundances (from Table 1) on a cosmochemical scale and an astronomical scale, and are compared with solar photospheric abundances and published meteorite abundances. The photospheric abundances for Li, Be and Cl are from Asplund et al. (2009), Na – Ca from Scott et al. (2015a), Sc – Ni from Scott et al. (2015b), and Cu – Th from Grevesse et al. (2015), and the meteoritic abundances for the CI chondrite Orgueil taken from Lodders et al. (2009). S.D. = standard deviations.

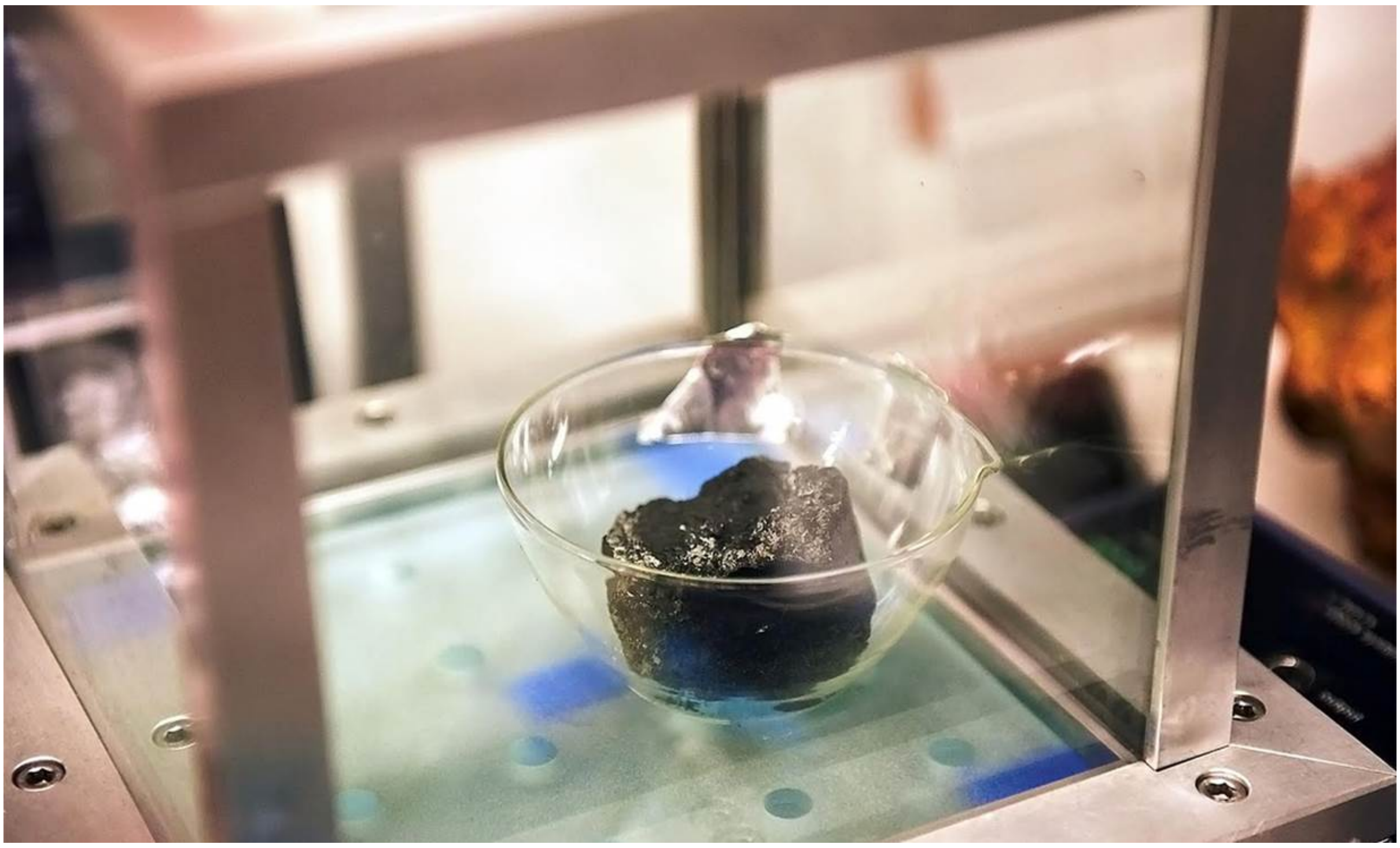

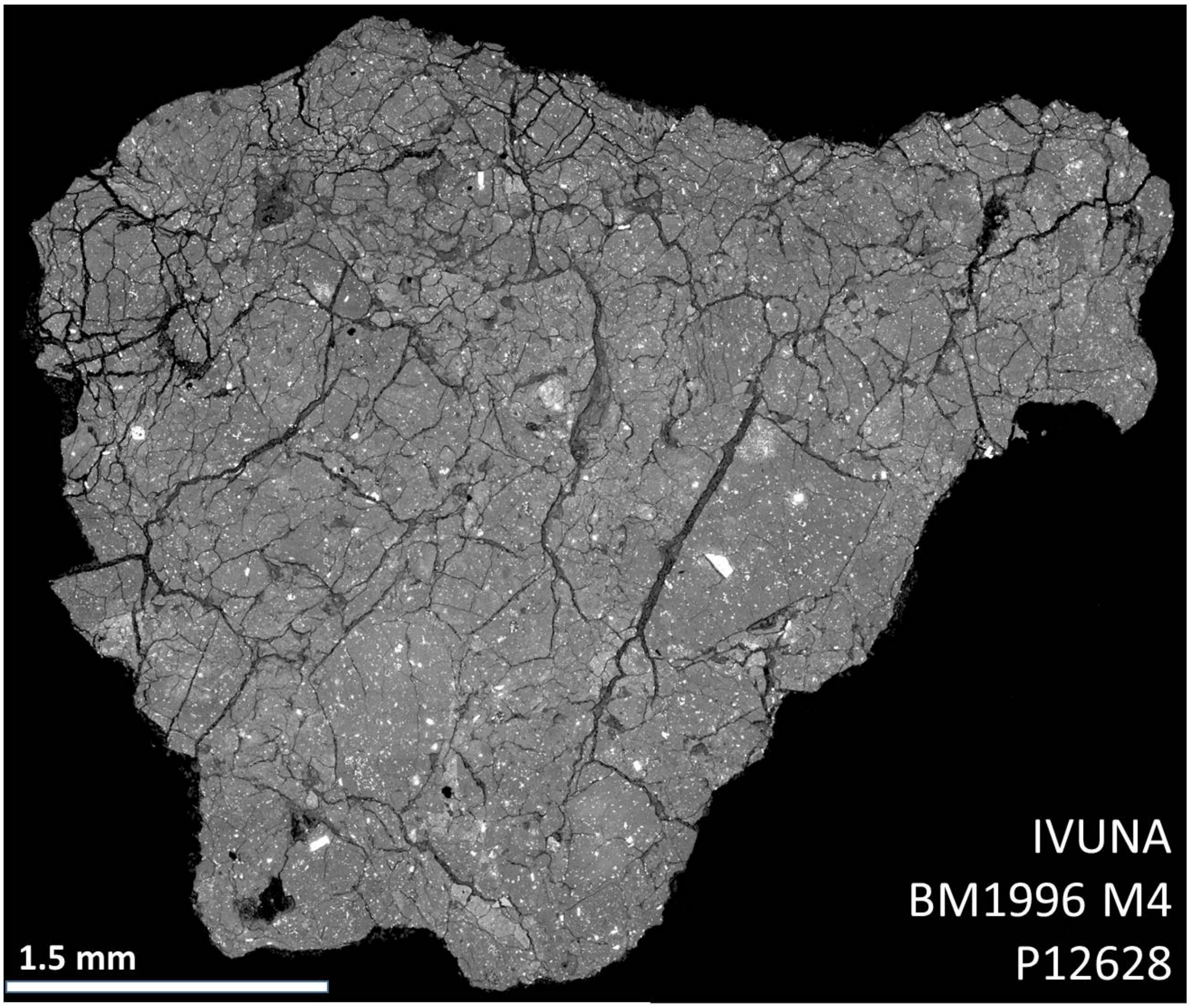
IVUNA
BM1996 M4
P12628
1.5 mm

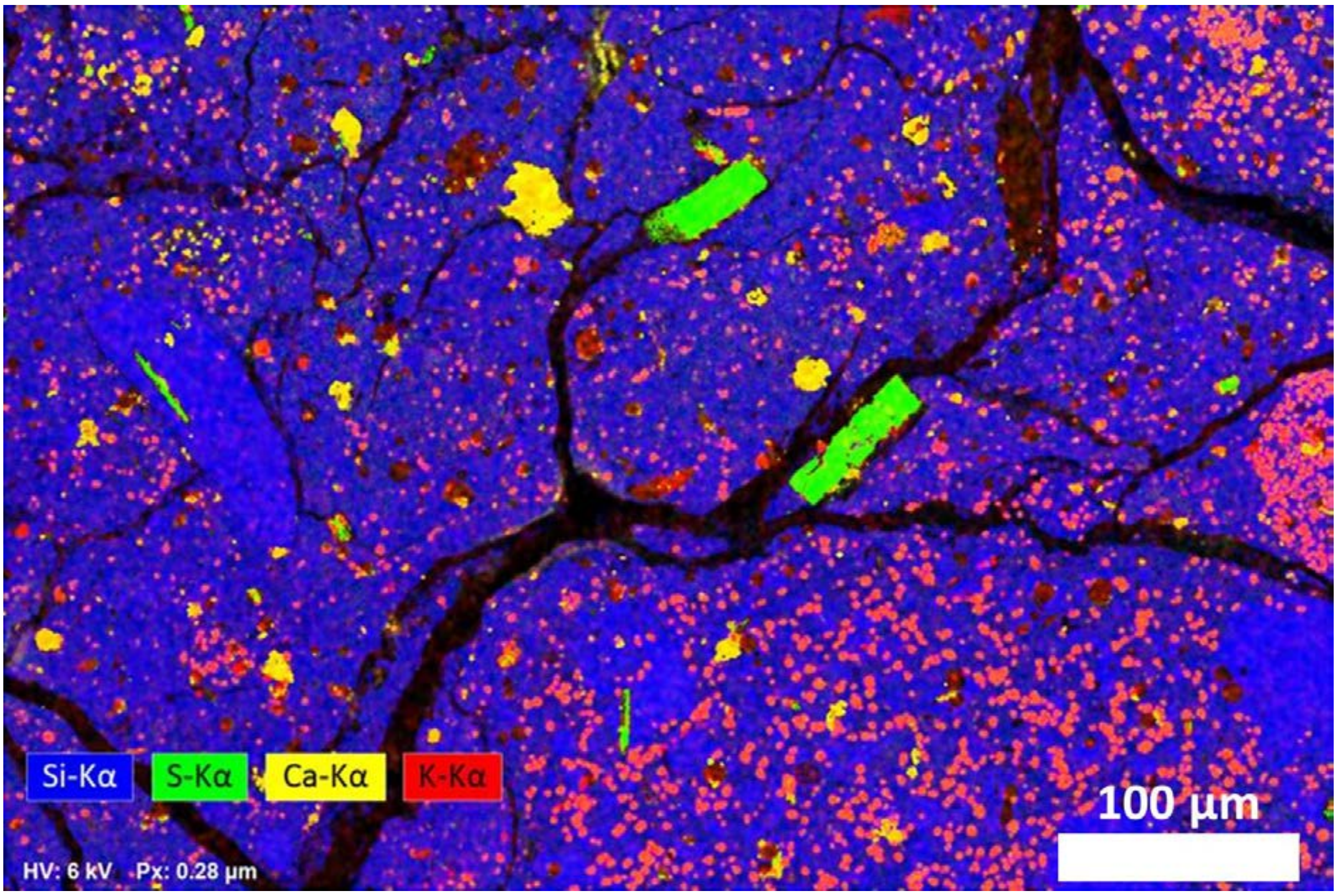

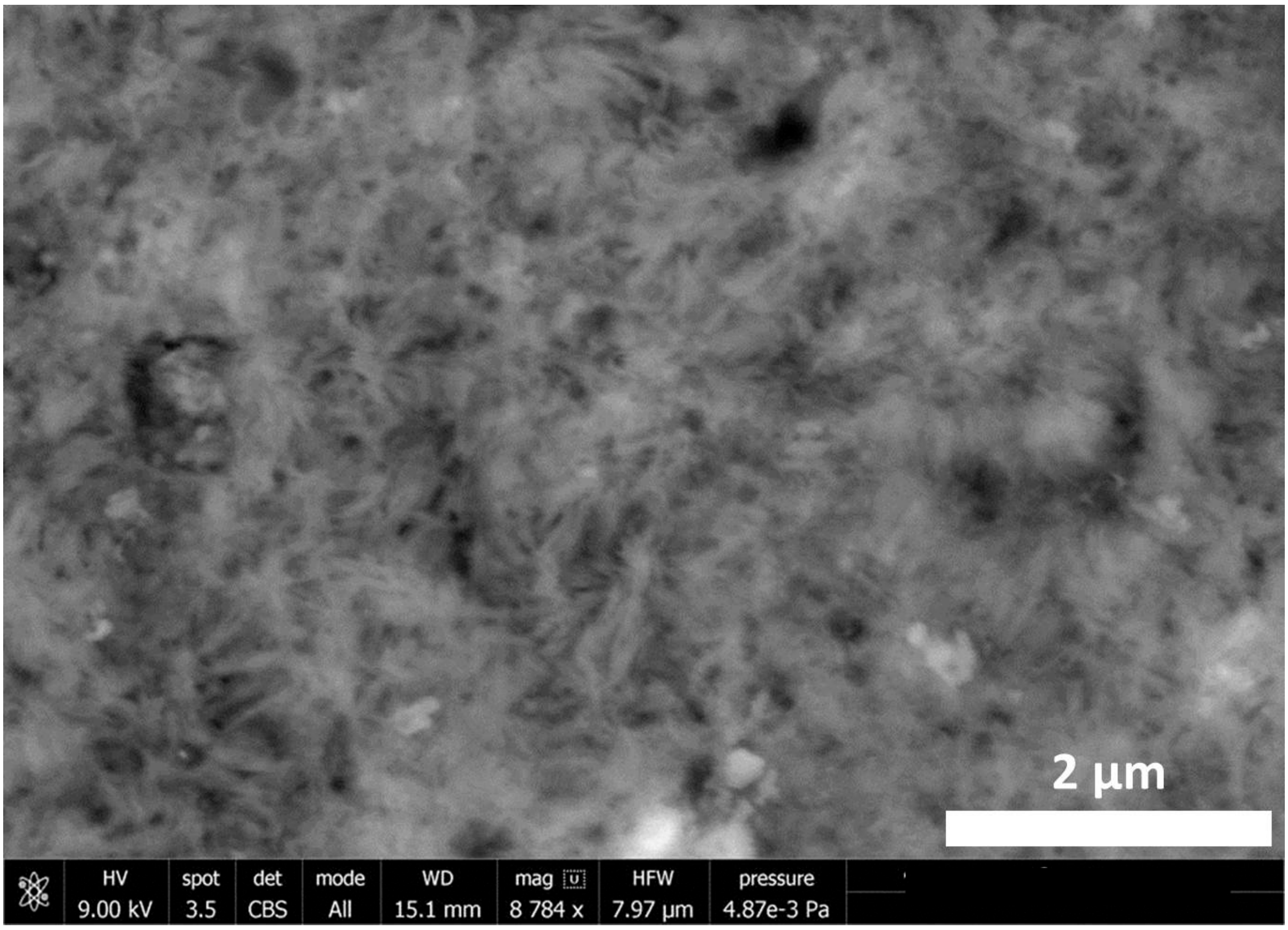

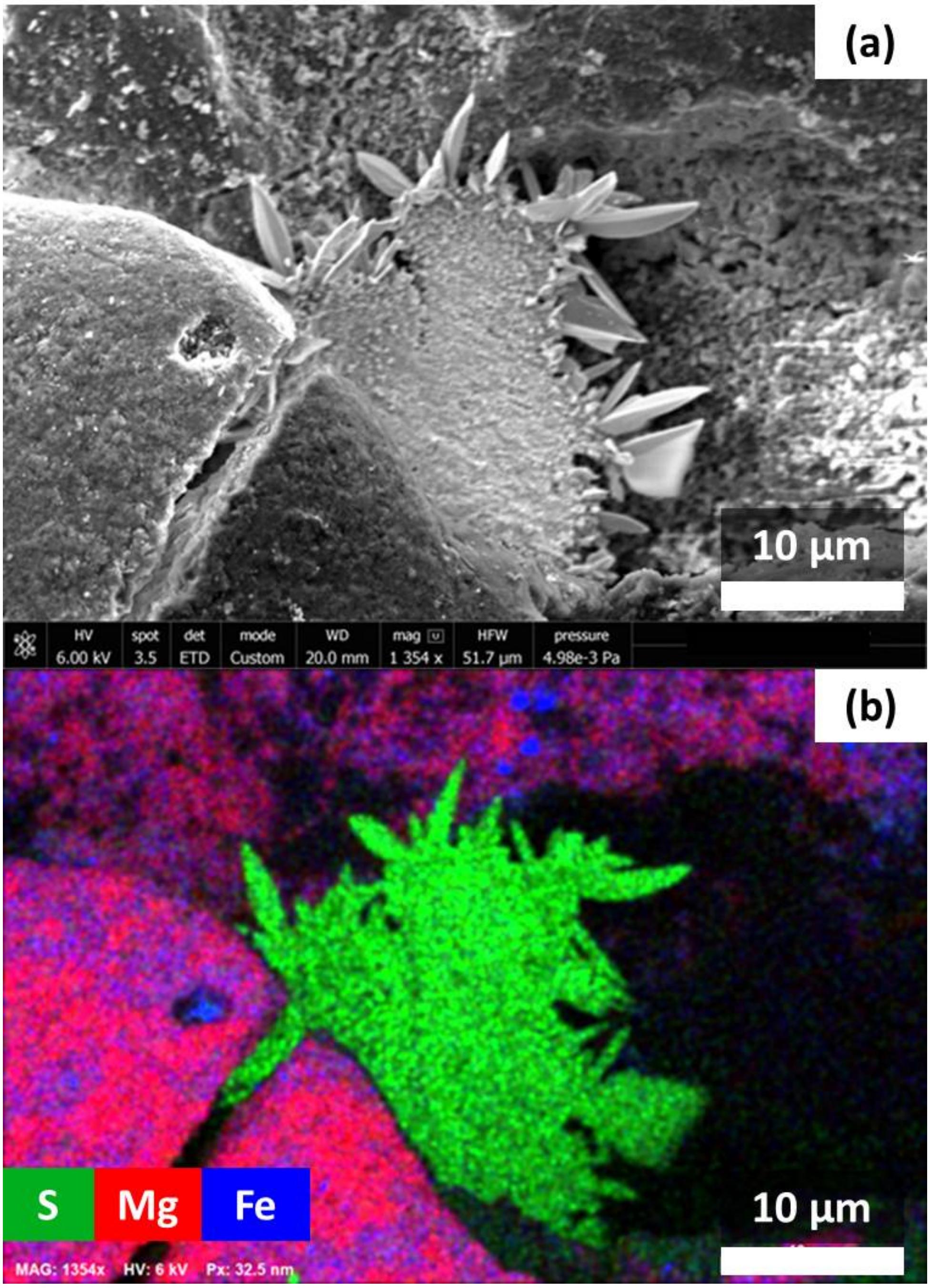
(a)
10 µm
HV 6.00 kV
spot 3.5
det ETD
mode Custom
WD 20.0 mm
mag 1 354 x
HFW 51.7 µm
pressure 4.98e-3 Pa
(b)
S
Mg
Fe
10 µm
MAG: 1354x HV: 6 kV Px: 32.5 nm

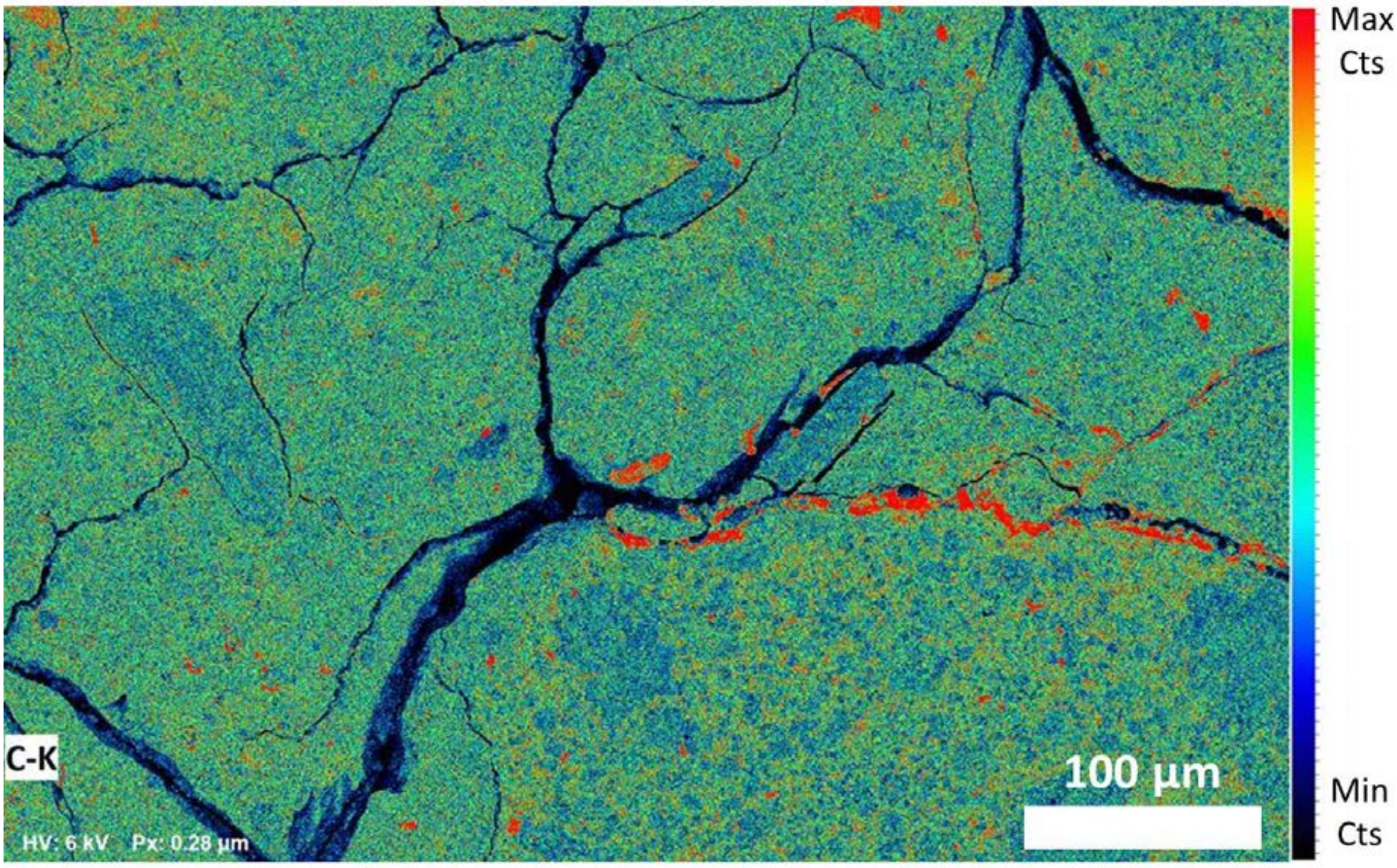

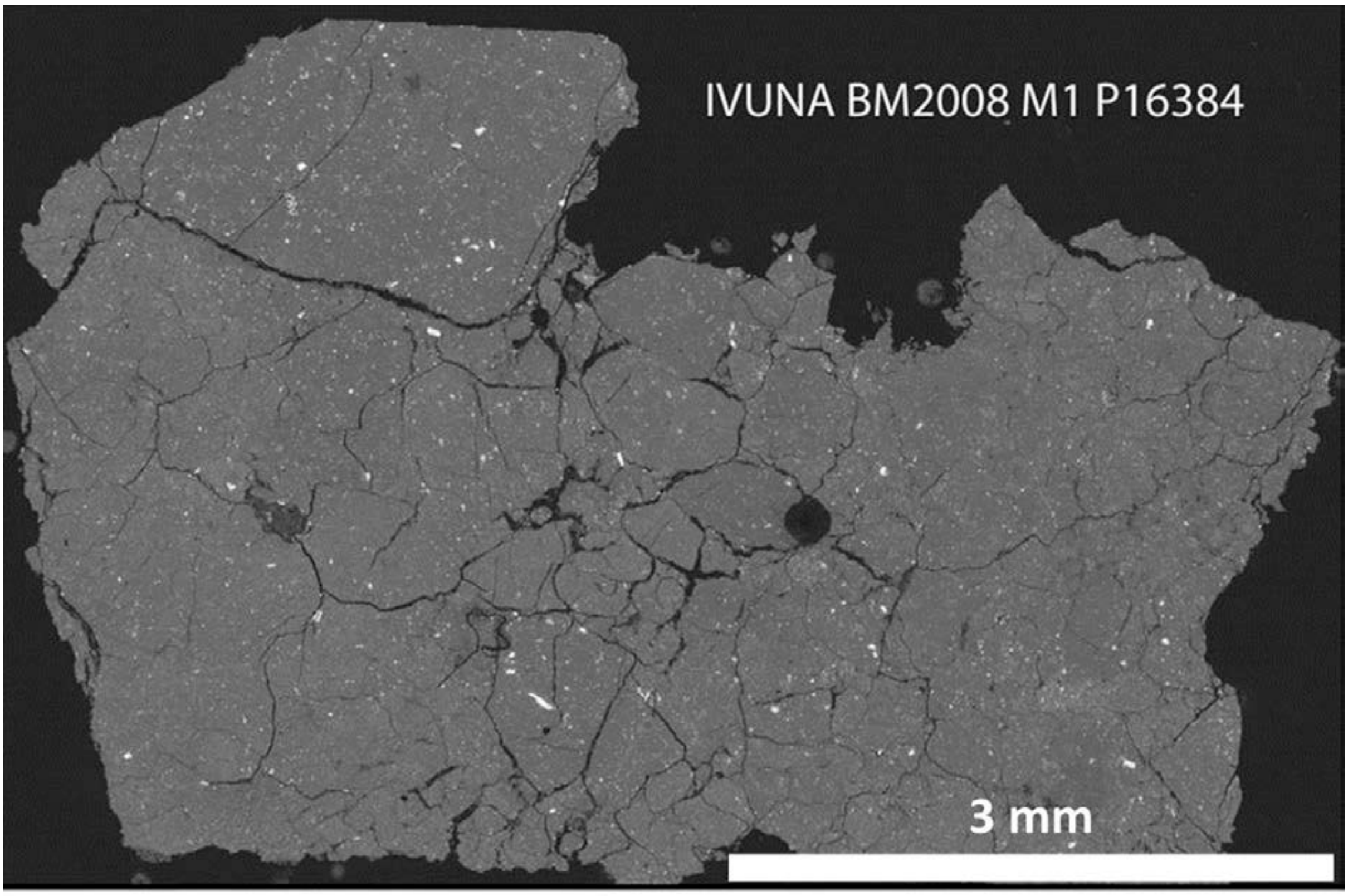
IVUNA BM2008 M1 P16384
3 mm

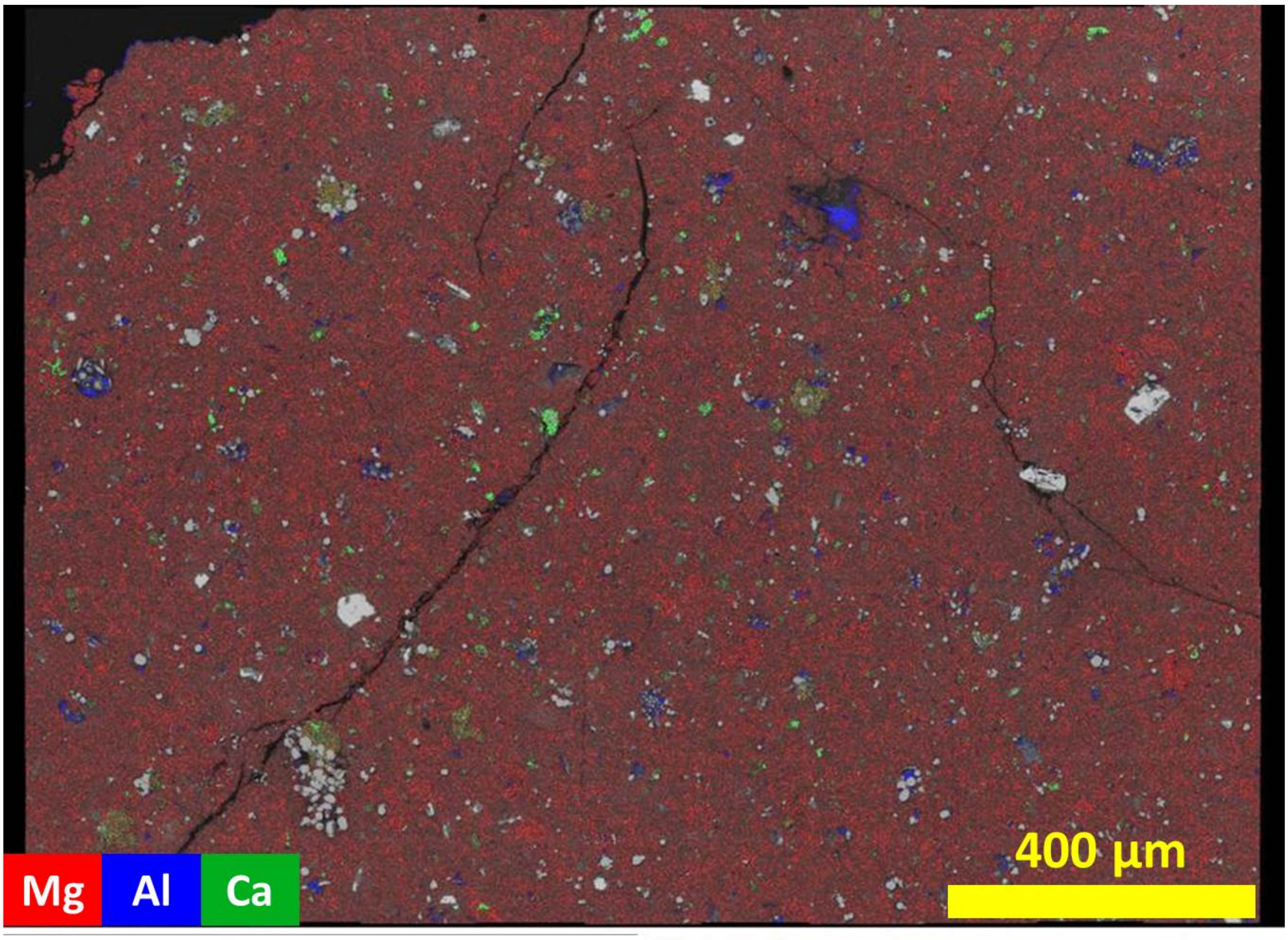
400 µm
Mg
Al
Ca

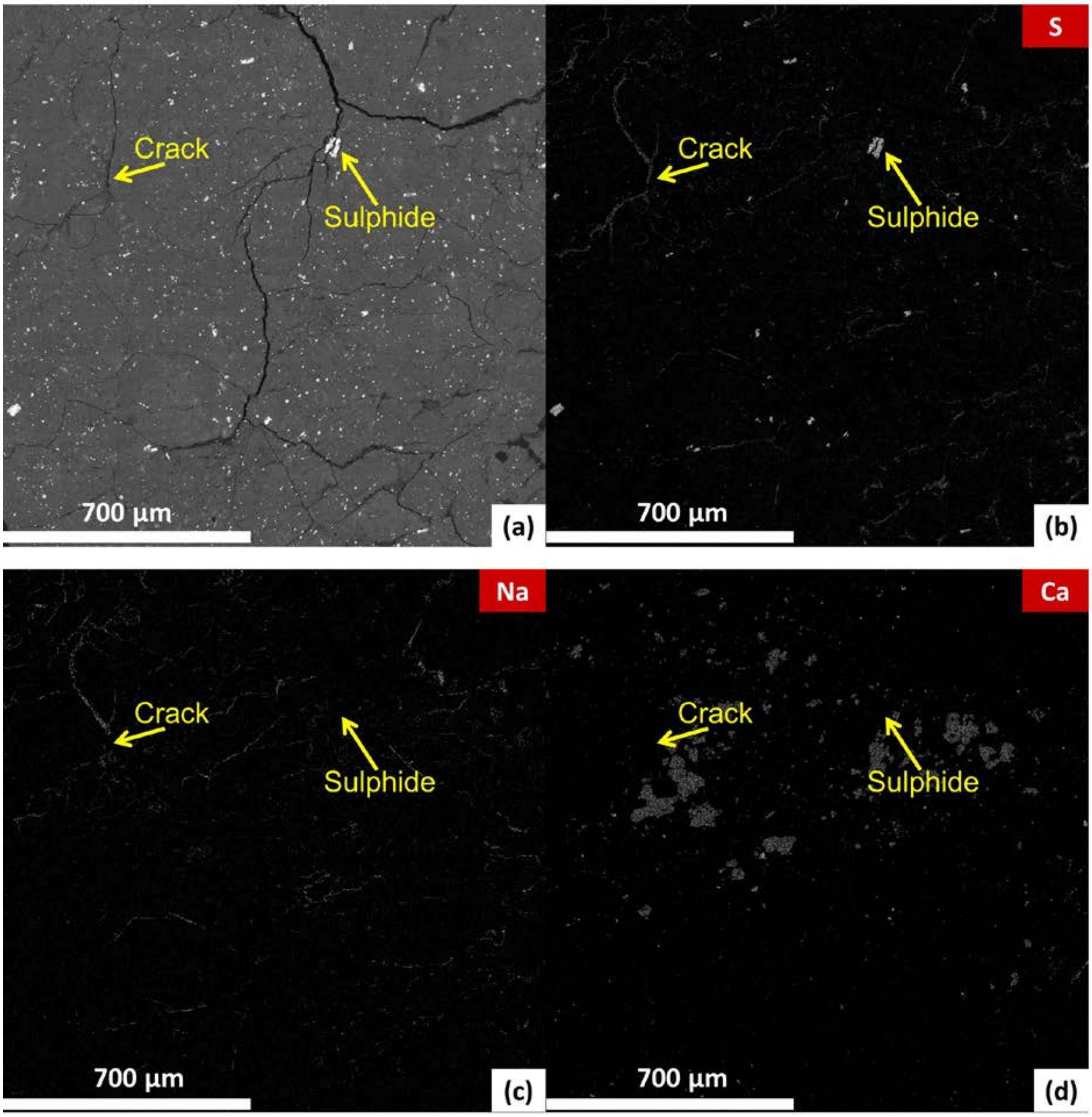

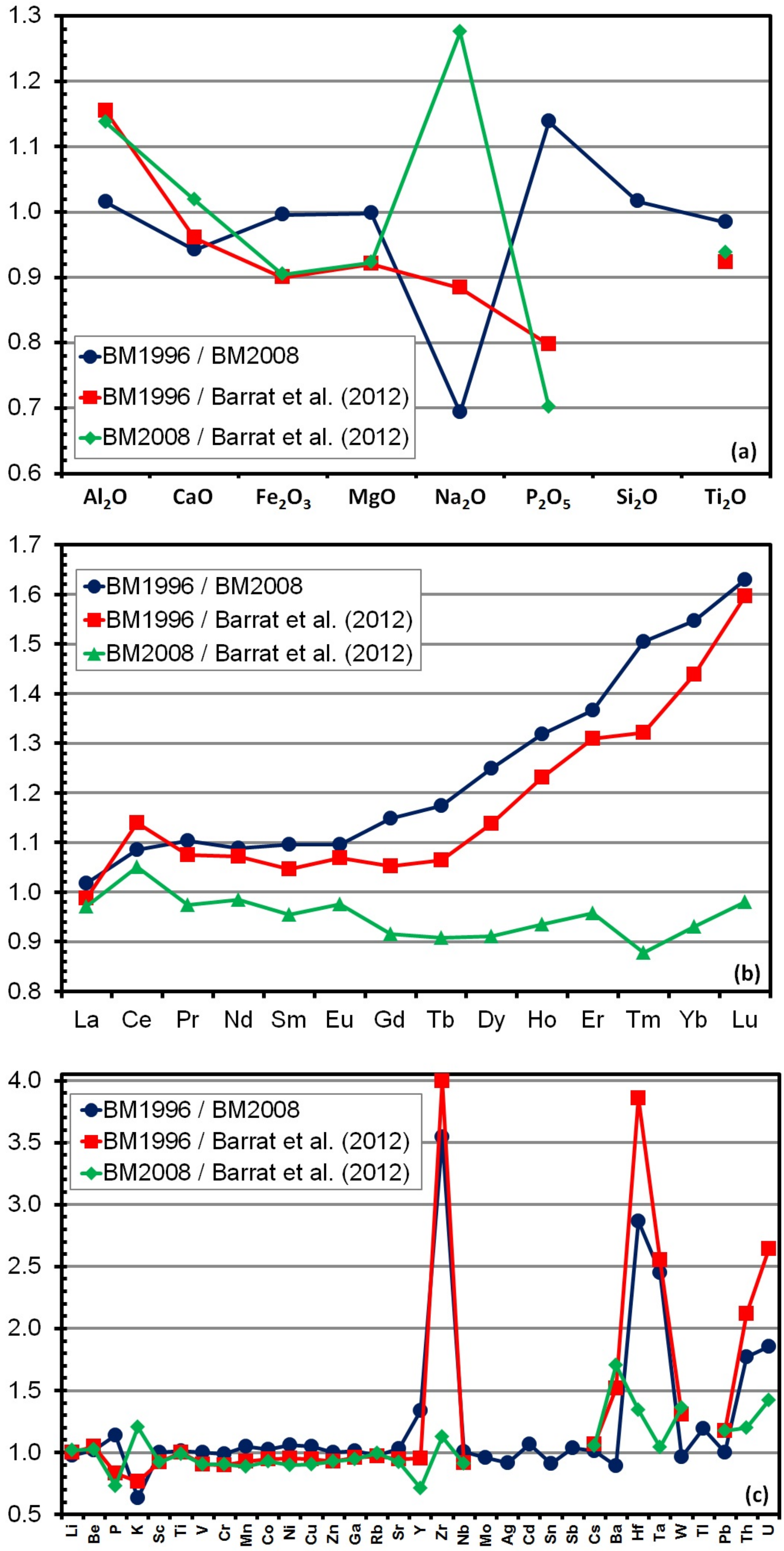

BM1996 / BM2008
BM1996 / Barrat et al. (2012)
BM2008 / Barrat et al. (2012)
(a)
Al2O CaO Fe2O3 MgO Na2O P2O5 Si2O Ti2O
(b)
La Ce Pr Nd Sm Eu Gd Tb Dy Ho Er Tm Yb Lu
(c)
Li Be P K Sc Ti V Cr Mn Co Ni Cu Zn Ga Rb Sr Y Zr Nb Mo Ag Cd Sn Sb Cs Ba Hf Ta W Tl Pb Th U

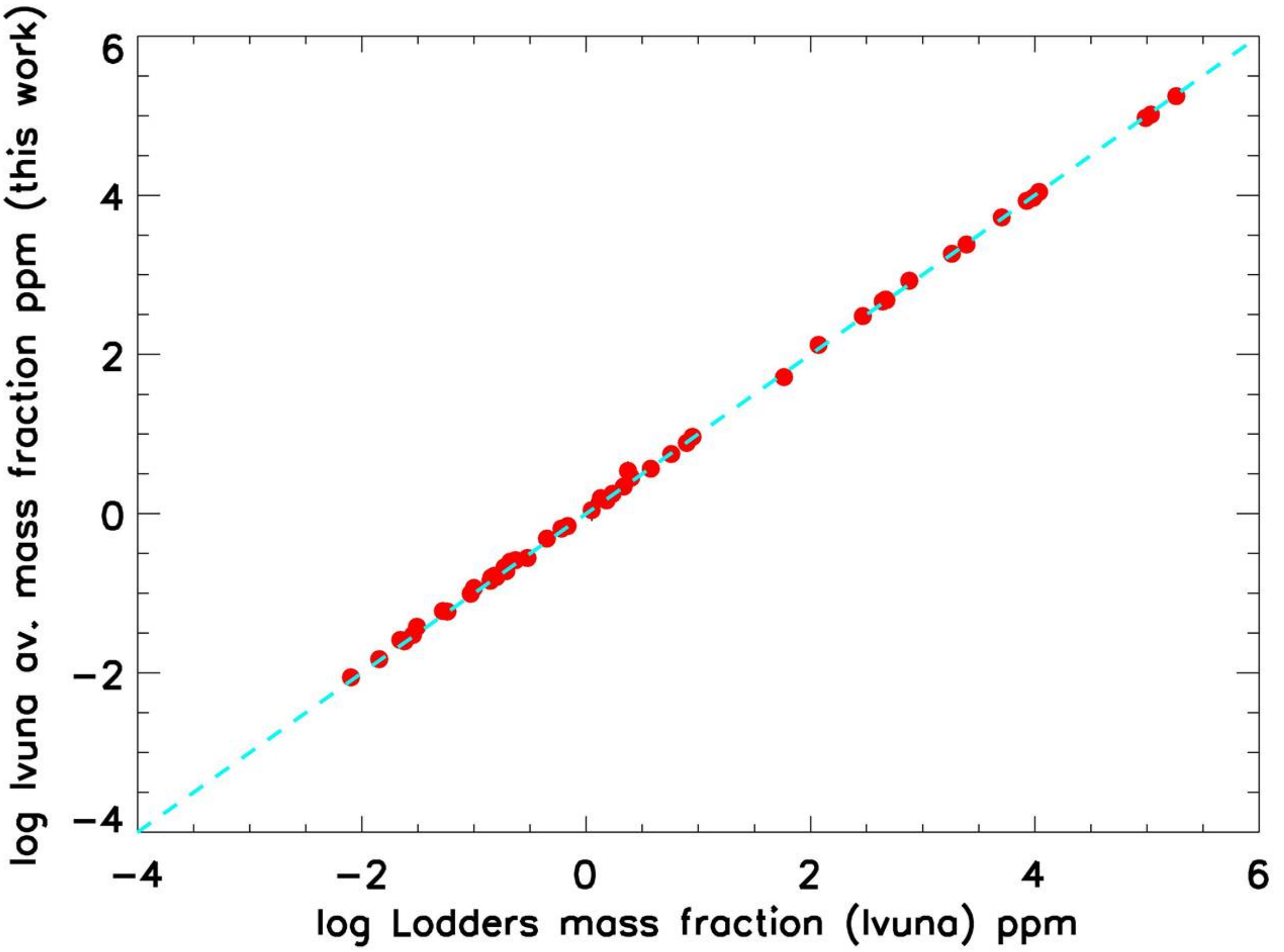
log Ivuna av. mass fraction ppm (this work)
log Lodders mass fraction (Ivuna) ppm
−4
−2
0
2
4
6

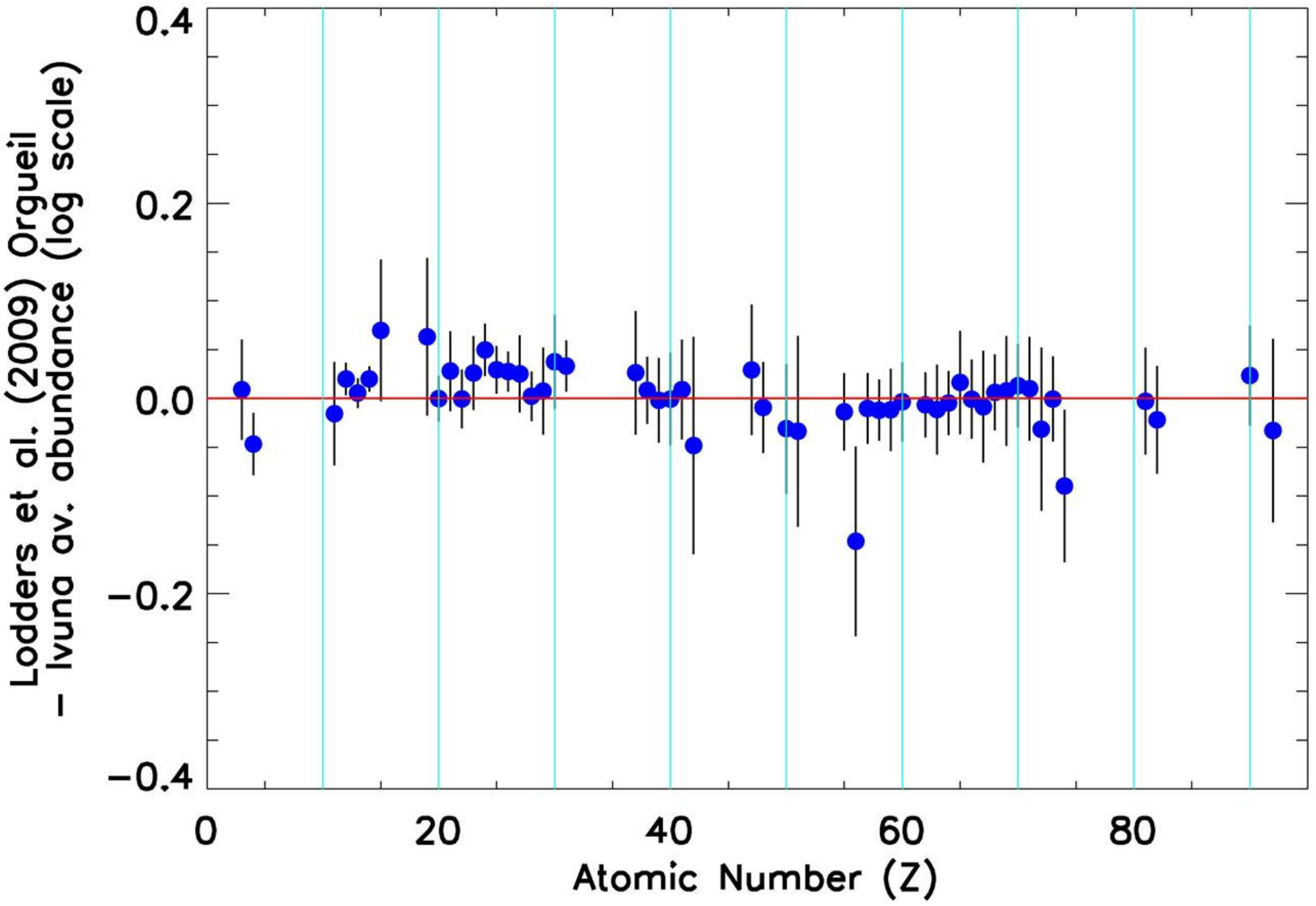

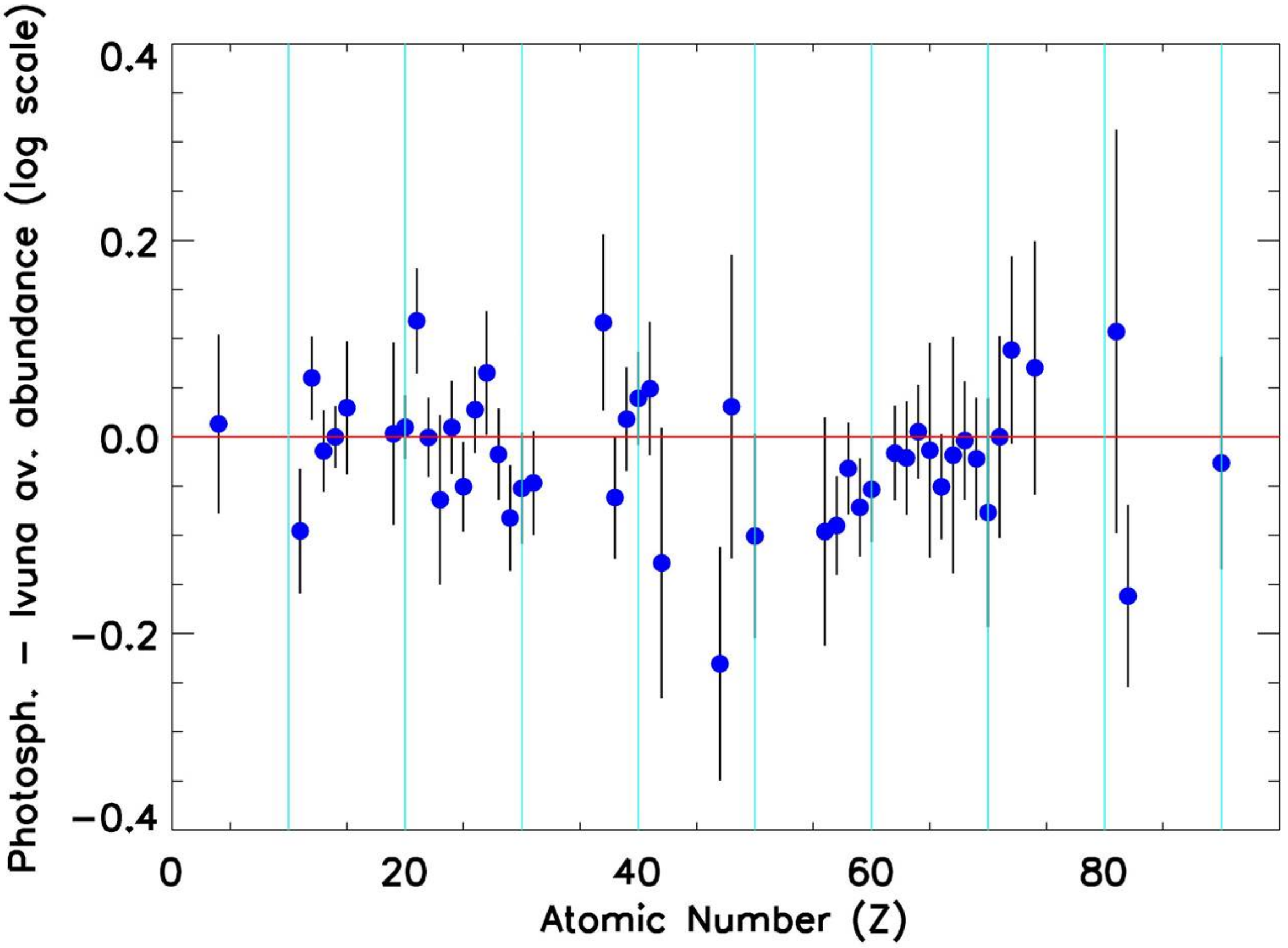